\documentclass[a4paper,12pt]{article}

\usepackage{graphicx}
\usepackage{citesort}
\usepackage{bm}

\begin{document}



\title{Neutrino mass in cosmology: status and prospects}


\author{Yvonne Y.~Y.~Wong \\
\small Institut f\"ur Theoretische Teilchenphysik und Komsologie, RWTH Aachen \\
\small D-52056 Aachen, Germany \\
\small yvonne.wong@physik.rwth-aachen.de}


\maketitle

\begin{abstract}
I give an overview of the effects of neutrino masses in cosmology,
focussing on the role they play  in
the evolution of cosmological perturbations.  
I discuss how recent observations of the cosmic microwave background anisotropies
and the large-scale matter distribution can probe neutrino masses
with greater precision than current laboratory experiments.  I 
describe several new techniques that will be used to probe cosmology in 
the future, as well as recent advances in the computation of the nonlinear 
matter power spectrum and related observables.
\end{abstract}


\newpage

\tableofcontents

\newpage

\section{Introduction}

Standard big bang theory predicts some  $10^{87}$ neutrinos per 
flavour in the visible universe, an abundance second only to the cosmic 
microwave background (CMB) photons.  Relativistic 
neutrinos constitute a significant fraction of the total energy density 
in the epoch of radiation domination, and play an important role in the 
primordial synthesis of light elements.
Even in today's matter and dark energy-dominated universe, a minimum neutrino 
mass of $0.07 \ {\rm eV}$ as established by the solar and atmospheric 
neutrino oscillations experiments \cite{nuexp} means that relic big bang neutrinos can 
still contribute  at least $0.5\%$ of the total matter density.  These 
neutrinos have a definite impact on cosmic structure formation.

This article provides a brief overview of the subject of neutrino cosmology, 
focussing  on how precision cosmological probes such as CMB anisotropies and the large-scale structure (LSS) distribution 
of matter can be used to determine or constrain the absolute
neutrino mass scale.  Readers interested in the role of neutrinos for big bang nucleosynthesis should
 consult the review of~\cite{bbn}.
After introducing the relic neutrino background in
section~\ref{sec:relic}, I outline the theoretical framework for computing precision cosmological
observables using linear perturbation theory in section~\ref{sec:linear}.  In section~\ref{sec:present} I describe the 
capacities of present observations to probe the absolute neutrino mass scale, 
while in sections~\ref{sec:nbody} and \ref{sec:analytic}
I report on some recent developments in the computation of nonlinear corrections to the LSS matter power spectrum and 
the halo mass function.   Section~\ref{sec:future} contains descriptions of 
various future observations and their forecasted sensitivities to the neutrino mass.  
I conclude in section~\ref{sec:conclusions}.

\section{The relic neutrino background\label{sec:relic}}
\subsection{Thermal history of the neutrino: decoupling}

In the first second after the big bang, the temperature of the universe is so high
that even weak interactions are able to hold neutrinos in thermal equilibrium with 
the cosmic plasma (photons, electrons, positrons, etc.). When in equilibrium,
the neutrinos' phase space density follows the Fermi--Dirac distribution,
\begin{equation}
\label{eq:fd}
f_\nu(E,T_\nu) = \frac{1}{1+\exp[(E-\mu)/T_\nu]},
\end{equation}
where the neutrinos share a common temperature with the cosmic plasma $T_\nu= T_\gamma \equiv T$,
and the chemical potential $\mu$ is expected, on theoretical grounds, to be similar to
the universal matter--antimatter asymmetry, $\mu/T_\nu \sim 10^{-10}$, i.e., negligible.
The number and energy densities per flavour are given respectively by
\begin{equation}
n_\nu = \frac{g}{(2 \pi)^3} \int d^3 p \ f_\nu(E,T_\nu), \qquad 
\rho_\nu = \frac{g}{(2 \pi)^3} \int d^3 p \ E \ f_\nu(E,T_\nu),
\label{eq:numden}
\end{equation}
where $g=2$ denotes the two helicity states, i.e.,
the expressions~(\ref{eq:numden}) count both neutrinos and antineutrinos.

As the universe expands and cools, the weak interaction rate drops off with temperature
as $\Gamma \sim G_F^2 T^5$.  When
interactions become too infrequent to match the Hubble expansion $H\sim T^2/m_{\rm planck}$,  
neutrinos decouple from the cosmic plasma.
By demanding $\Gamma \sim H$, we see 
that neutrinos decouple  at  $T \sim 1 \ {\rm MeV}$.  
Laboratory experiments indicate that the absolute neutrino mass should not exceed
${\cal O}( 1)$~eV~\cite{Lobashev:2003kt,Kraus:2004zw}.
 Thus, at decoupling, the neutrinos are relativistic, and their phase space density is well-approximated by the 
relativistic Fermi--Dirac distribution,
\begin{equation}
\label{eq:relfd}
f_\nu(E,T_\nu) = \frac{1}{1+\exp[p/T_\nu]},
\end{equation}
where for simplicity the chemical potential $\mu$ has been set to zero.

\subsection{After decoupling: the neutrino phase space and temperature}

Because neutrinos decouple while they are relativistic, 
they preserve, to an excellent approximation, their relativistic 
Fermi--Dirac phase space density even after they become 
nonrelativistic at later times.  The neutrino phase space after decoupling 
is a frequent point of confusion and deserves a detailed explanation.

Consider neutrinos with mass $m_\nu = 1$~eV decoupling at a temperature of 1~MeV.  By
integrating the Fermi--Dirac distribution~(\ref{eq:fd}), one finds that only 1 
in $10^{19}$ neutrinos has a momentum $p<m_\nu$.  This justifies the use of equation~(\ref{eq:relfd}) 
to describe the neutrino phase space distribution up to the point of decoupling.

Physical momentum $p$ redshifts as $a^{-1}$, where $a$ is the scale factor, 
for both  massless and massive particles.  It is thus convenient to 
define a non-redshifting momentum variable $y=a p$, so that the
neutrino number density~(\ref{eq:numden}) can be written as
$n_\nu  \propto a^{-3} \int dy^3 f_\nu(E,T_\nu)$.
After decoupling, particle number conservation requires that the $n_\nu$ scales as
$n_\nu \propto a^{-3}$.
Thus comparing these two expressions one sees immediately
that  agreement can only be achieved if the phase space density $f_\nu(E,T_\nu)$ 
is independent of $a$ at all times. Inspection of equation~(\ref{eq:relfd}) shows that 
this can be brought about if, similar to the physical momentum $p$, the neutrino temperature $T_\nu$ also 
redshifts as $a^{-1}$.  
Thus
the relativistic Fermi--Dirac distribution~(\ref{eq:relfd}) remains an apt description of the neutrino 
phase space density, even after the neutrinos  become nonrelativistic.

Immediately after decoupling the neutrinos have the same temperature as the cosmic plasma.
This correspondence remains true for some time because both $T_\nu$ and $T_\gamma$  scale as $a^{-1}$.
At $T \sim m_e/3 \sim 0.2$~MeV, electrons and positrons become nonrelativistic.  Their annihilation 
into photons leads to the transfer of entropy originally residing in the $e^+ e^-$ fluid to the photon fluid,
which causes the latter to heat up.  This reheating is, however, not felt by the decoupled neutrinos. 
 Hence the neutrinos will emerge from the $e^+ e^-$ annihilation
event colder than the photons.   One can show using entropy conservation arguments (see, e.g., \cite{Lesgourgues:2006nd}) that
the new neutrino temperature is related to the photon temperature via
\begin{equation}
T_\nu = \left(\frac{4}{11}\right)^{1/3} T_\gamma. \label{eq:tnu}
\end{equation}
This expression remains valid  today.

Equation~(\ref{eq:tnu}) provides a useful estimate for the temperature of the 
relic neutrino background.  However, we do expect some small corrections owing to 
the fact that the processes of neutrino decoupling and $e^+ e^-$ annihilation 
occur in close proximity in time and that these processes are not instantaneous;
At the time of $e^+ e^-$ annihilation, some neutrinos, particularly those at the 
high energy tail of the Fermi--Dirac distribution, are still coupled to the 
cosmic plasma and will partake in the reheating process.  
Thus in general we expect the neutrino energy density to be a little higher
than is implied by the relation~(\ref{eq:tnu}).  

This increase in the neutrino energy density is usually parameterised
in terms of an increase in the  effective number of neutrino families $N_{\rm eff}$,
defined via $\sum_i \rho_{\nu,i} \equiv N_{\rm eff} \times \rho_{\nu,0}$,
where  $\sum_i \rho_{\nu,i}$ is the total energy density residing in all neutrino species,
and $\rho_{\nu, 0} = (7/8)  (\pi^2/30) g T_\nu^4$
denotes the ``standard'' neutrino energy density per flavour.  Evidently,  this relation  is uniquely
defined only at early times when the neutrinos are still relativistic.
Taking also into account  neutrino  flavour oscillations and
 finite temperature QED effects, $N_{\rm eff} = 3.046$~\cite{neff}.

\subsection{Properties of the relic neutrino background today}

Using equation~(\ref{eq:tnu}) and measurements of the present CMB temperature
$T_{{\rm CMB},0} = 2.725 \pm 0.001$~K~\cite{Fixsen:1996nj} leads us to expect 
 a relic neutrino background of temperature 
$T_{\nu,0} = 1.95 \ {\rm K} \sim 10^{-4} \ {\rm eV}$.
The number density is expected to be $112 \ {\rm cm}^{-3}$ per flavour from equation~(\ref{eq:numden}).

The exact energy density per flavour depends on whether the neutrinos are relativistic or nonrelativistic
today.   The neutrinos are relativistic if $m_\nu \ll T_{\nu,0}$, in which case their 
energy density per flavour is $\rho_\nu = (7/8)  (4/11)^{4/3} \rho_{\rm CMB}$.
Normalised to the present-day critical density $\rho_{{\rm crit},0} = 3 H_0^2/8 \pi G$, where $H_0=100 \ h$ is the present
Hubble parameter and $G$ Newton's constant, we find
$\Omega_\nu h^2 \equiv  (\rho_{\nu,0}/\rho_{{\rm crit},0}) h^2  = 6 \times 10^{-6}$.
In other words, the energy density due to relativistic neutrinos today is completely negligible.

However, if $m_\nu \gg T_{\nu,0}$,
then the energy density per flavour is $\rho_\nu = m_\nu n_\nu$,
or, equivalently,
\begin{equation}
\label{eq:omeganu}
\Omega_\nu h^2 \simeq \frac{m_\nu}{93 \ {\rm eV}}.
\end{equation}
Thus, even for a neutrino mass as small as $m_\nu=0.05 \ {\rm eV}$, we expect to find a non-negligible 
$\Omega_\nu \sim 0.1$\%;  These neutrinos then form a  dark matter component in the universe.
By demanding that massive neutrinos not overclose the universe, 
i.e., $\Omega_\nu < 1$, one can immediately set an upper bound on the sum of the neutrino masses, 
$\sum m_\nu < 93 \ {\rm eV}$
\cite{Gershtein:1966gg,Cowsik:1972gh}.
Historically this is first upper bound on the neutrino mass  from cosmology and is sometimes known 
as the ``closure'' bound.

\section{Linear cosmological perturbations\label{sec:linear}}

Neutrino dark matter satisfying the closure bound
cannot constitute all of the dark matter content of the universe because thermal relics that decouple 
when relativistic come with a large velocity dispersion.  Using the relativistic Fermi--Dirac 
distribution~(\ref{eq:relfd}), we find a rough estimate of this velocity dispersion:
\begin{equation}
\label{eq:thermal}
\langle v_{\rm thermal} \rangle \simeq 81 (1+z) \left( \frac{\rm eV}{m_\nu} \right) \, {\rm km} \ {\rm s}^{-1}.
\end{equation}
For a $\sim 1$~eV neutrino, $\langle v_{\rm thermal} \rangle \simeq 100 \, {\rm km} \ {\rm s}^{-1}$ is 
comparable to the typical velocity dispersion of a galaxy.  For dwarf galaxies, the velocity dispersion is 
even smaller, $\sim O(10) \, {\rm km} \ {\rm s}^{-1}$.  Thus the relic neutrinos  have much too much thermal
energy  to be squeezed into small volumes to form the smaller structures we observe today~\cite{tremainegunn}.  
In contrast, cold dark matter (CDM) has by definition 
$\langle v_{\rm thermal} \rangle=0$, and is thus not subject to these constraints.

Nonetheless,  even if  relic neutrinos cannot form the bulk of the cosmic dark matter, because their kinematic properties 
are so different from those of CDM, their presence at even the $\Omega_\nu \sim 0.1$\% level 
must leave a signature in the large-scale cosmological observables.
Detecting this signature will then allow us to establish the absolute neutrino mass scale via equation~(\ref{eq:omeganu}).

In this section, I outline the theoretical framework for predicting the effects of massive neutrinos on the CMB anisotropies 
and LSS matter power spectrum
via linear perturbation theory.  For more detailed discussions of linear cosmological perturbation theory in general, see, 
e.g.,~\cite{mabertschinger,durrer}.    

\subsection{The homogeneous universe}

The observed universe appears to be homogeneous and isotropic on scales  of ${\cal O}(100)$~Mpc.   
On these scales space also appears to be expanding.   
The simplest spacetime metric that captures these observational features has the form
\begin{equation}
\label{eq:flrw}
ds^2 =g_{\mu\nu} dx^\mu dx^\nu= a^2(\tau) [ - d\tau^2 + \gamma_{ij} dx^i dx^j],
\end{equation}
where $\tau$ is the conformal time, and $x^i \doteq {\mathbf x}$ are the comoving coordinates.  
The metric~(\ref{eq:flrw}), known as the Friedmann--Lema\^{\i}tre--Robertson--Walker (FLRW) metric, 
forms the basis of modern cosmology.

The spatial part of the FLRW metric $\gamma_{ij}$ encodes the local geometry of space, which can be  
 (i)  flat and Euclidean, (ii) spherical (i.e., with positive curvature), or (iii)
hyperboloid (i.e., with negative curvature).   Currently, 
there is no observational evidence for spatial curvature~\cite{Komatsu:2010fb}.  
From a theoretical perspective, it is also difficult to reconcile spatial curvature with inflationary cosmology 
(see, e.g., \cite{riotto}).    We therefore consider only the case of flat spatial geometry, so that
$\gamma_{ij} = \delta_{ij}$.

The energy content of the universe is encoded in the stress--energy tensor $T_{\mu\nu}$.  Homogeneity and
 isotropy imply that there is only one sensible choice,
\begin{equation}
\label{eq:tmunu}
T^{\mu}_{\ \nu} = \bar{T}^\mu_{\ \nu}  \equiv {\rm diag}\left(- \bar{\rho}, \ \bar{p},\ \bar{p}\ ,\bar{p} \right),
\end{equation}
where $\bar{\rho}$ and $\bar{p}$ are the spatially averaged energy density and pressure, respectively, of a 
comoving fluid in its rest frame.  Expression~(\ref{eq:tmunu})
can be easily generalised to the multi-fluid case with the replacements  
$\bar{\rho}  \to \sum_\alpha \bar{\rho}_\alpha$ and  $\bar{p}  \to \sum_\alpha \bar{p}_\alpha$, where 
$\alpha$ sums over all fluids present:  this usually means CDM $\alpha=c$, neutrinos $\nu$, baryons~$b$ 
(i.e., protons and electrons),
 and photons $\gamma$.

Inserting the FLRW metric~(\ref{eq:flrw}) and the stress--energy tensor~(\ref{eq:tmunu}) into the Einstein equation,
$R_{\mu\nu}- \frac{1}{2} g_{\mu \nu} R = 8 \pi G T_{\mu \nu}$,
we find the Friedmann equations,
\begin{equation}
{\cal H}^2 \equiv \left( \frac{\dot{a}}{a} \right)^2 = \frac{8 \pi G}{3} a^2 \bar{\rho}, \qquad
\dot{\cal H} = - \frac{4 \pi G}{3} a^2 (\bar{\rho} + 3 \bar{p}),
\end{equation}
with $\cdot  \equiv d/d \tau$.  Likewise, conservation of energy--momentum, $\nabla_\mu T^{\mu \nu}=0$, implies
$\dot{\bar{\rho}} + 3 {\cal H} (\bar{\rho} + \bar{p}) = 0$,
from which we deduce that $\bar{\rho} \propto a^{-3}$ for nonrelativistic matter, $\propto a^{-4}$ for radiation, 
and so forth.

\subsection{Metric perturbations and the perturbed Einstein equation}

Let us now consider the case of small perturbations around the FLRW 
spacetime.
In its most general form the perturbed metric can be written as
\begin{equation}
\label{eq:general}
ds^2 = a^2(\tau)\{  -(1+2 A) d\tau^2 - 2 B_i d\tau dx^i + [(1+ 2 H_L) \delta_{ij} + 2 h_{ij}] dx^i dx^j\}.
\end{equation}
Here, the quantities $A$ and $H_L$ represent two scalar degrees of freedom (d.o.f.) under rotation; 
$B_i$ is a vector field which can be further decomposed into 
a curl-free scalar part $B^{(s)}_i$ (1 d.o.f.) and a divergence-free vector part $B^{(v)}_i$ (2 d.o.f.); 
$h_{ij}$ is a traceless $3 \times 3$ matrix decomposable into a 
scalar/longitudinal
$h^{(s)}_{ij}$ (1 d.o.f.), a vector/solenoidal $h^{(v)}_{ij}$ (2 d.o.f.) and a tensor/transverse $h^{(t)}_{ij}$ (2 d.o.f) component. 
Starting with the general expression for the stress--energy tensor,
\begin{equation}
T^{\mu\nu} =  \bar{T}^{\mu \nu}+ \delta T^{\mu\nu}  = (\rho + p) u^\mu u^\nu + pg^{\mu \nu} + \Sigma^{\mu \nu},
\end{equation}
where $u^\mu = d x^\mu/(-d s^2)^{1/2}$ is the fluid's 4-velocity,  $\rho$ and $p$ the rest frame energy density and pressure respectively,
and $\Sigma$  the shear stress, we can likewise decompose $ \delta T_{\mu \nu}$ under the rotation group.
The advantage of this decomposition is that once we insert the perturbed metric and stress--energy tensor into the
 Einstein equation, the resulting evolution equations for the scalar, vector and tensor perturbed components  are 
 completely decoupled from one another at linear order.

Of particular interest to us are the scalar modes.
 The scalar part of the perturbed stress--energy tensor takes the form
 \begin{eqnarray}
\delta T^0_{\ 0} = -  \delta \rho, & \qquad &  \delta T^i_{\ 0} = -(\bar{\rho}+\bar{p}) {v^{(s)}}^i, \nonumber \\
\delta T^0_{\ i} =  (\bar{\rho}+\bar{p}) v^{(s)}_i- B^{(s)}_i,  &\qquad&
\delta T^i_{\ j} = \delta p \ \delta^i_j +{ \Sigma^{(s)}}^ i_{\, j},
\end{eqnarray}
where $\rho \equiv \bar{\rho} + \delta \rho \equiv \bar{\rho}(1 + \delta)$ and similarly for $p$, and 
$v^{(s)}_i$ denotes the scalar component of the coordinate velocity $v_i \equiv d x_i/d \tau$.
The decoupling of the rotational modes at the linear level  means that we are free to consider only the scalar part
 of the metric perturbations, i.e.,
instead of handling simultaneously 10 d.o.f.\ of the general perturbed metric~(\ref{eq:general}),
we  need only to deal with the 4 scalar d.o.f., $A$, $H_L$, $B^{(s)}_i$ and $h^{(s)}_{ij}$.
Of these 4 d.o.f., only two are physical; the others  are gauge modes related to our choice of the 
background coordinates.  In other words, we are free 
to set any 2 d.o.f.\ to zero. Doing so is termed fixing or choosing a gauge. 

A number of popular gauges are used today in cosmological perturbation theory, each with its own advantages 
(e.g., physical intuition, numerical stability, etc.).  I will use the  conformal Newtonian gauge, which consists of setting 
$B^{(s)}_i=0$ and $h^{(s)}_{ij}=0$, so that the scalar perturbed FLRW metric now reads
\begin{equation}
\label{eq:conformal}
ds^2 = a^2(\tau)\{-[1 + 2 \psi({\mathbf x},\tau)] d\tau^2 + [1-2 \phi({\mathbf x},\tau)] \delta_{ij} dx^i dx^j \},
\end{equation}
where we have also relabelled $A \equiv \psi$ and $H_L \equiv - \phi$.

One further decomposition of convenience is the Fourier decomposition, given that in linear theory each eigenmode of the Laplacian evolves independently from each other.
Define the Fourier transform as
\begin{equation}
\tilde{A}({\mathbf k},\tau) ={\cal F}[A({\mathbf x},\tau) ] =  \int \frac{d^3 x}{(2 \pi)^3}  A({\mathbf x},\tau) e^{-i {\mathbf k} \cdot {\mathbf x}}.
\end{equation}
We can now proceed to write down the perturbed Einstein equation for the scalar modes in Fourier space to 
linear order in the perturbations:
\begin{eqnarray}
\label{eq:perturbedee}
-k^2 \tilde\phi - 3 {\cal H} (\dot{\tilde{\phi}} + {\cal H} \tilde{\psi}) &=& 4 \pi G a^2 \bar{\rho} \tilde{\delta},  \nonumber \\
\dot{\tilde\phi} + {\cal H} \tilde\psi&=&4 \pi G a^2 (\bar{\rho}+\bar{p}) \tilde\theta, \nonumber \\
\ddot{\tilde\phi} + {\cal H} (\dot{\tilde\psi} + 2 \dot{\tilde\phi}) + (2 \dot{\cal H}+ {\cal H}^2) \tilde\psi + \frac{k^2}{3} (\tilde\phi-\tilde\psi) &= &4 \pi G a^2 \tilde{\delta p}, \nonumber \\ 
 k^2 (\tilde\phi-\tilde\psi)&=& 12 \pi G a^2 (\bar{\rho}+\bar{p}) \tilde\sigma, 
\end{eqnarray}
where $k \equiv |{\mathbf k}|$,  $\tilde\theta  \equiv  i k^i \tilde{v}_i$ is the velocity divergence, and 
$(\bar{\rho} + \bar{p}) \tilde\sigma \equiv  - (k^{-2} k_i k_j-\frac{1}{3} \delta_{ij}) \tilde\Sigma^{ij}$ the anisotropic stress.
Note that these constructions naturally project out the scalar components of $v_i$ and $\Sigma^i_{\ j}$.

\subsection{Boltzmann equation}

Now it remains to write down the evolution equations for the quantities appearing on the right hand side of the perturbed Einstein equation~(\ref{eq:perturbedee}).  The starting point is the phase space density of a fluid $f(x^i, P_j,\tau)$, defined so that
$dN=g/(2 \pi)^3 f(x^i, P_j,\tau) dx^1 dx^2 dx^3 dP_1 dP_2 dP_3$
gives the number of particles in a phase space volume $dx^1 dx^2 dx^3 dP_1 dP_2 dP_3$.  
Here, $P^i$ is the 4-momentum of a fluid element,  related
to the proper momentum $p^i \doteq {\mathbf p}$ measured by a comoving observer in  the coordinates~(\ref{eq:conformal}) via
$P_i = a (1-\phi) p_i$.  It is also useful
to define a new momentum variable ${\mathbf q} \equiv a {\mathbf p}$, since in the case $\phi=0$, $q \equiv |{\mathbf q}|$ is non-redshifting.

The general covariant expression for the stress--energy tensor in terms of the phase space density of a fluid  $f(x^i, P_j,\tau)$ is
\begin{equation}
T_{\mu \nu} =\frac{g}{(2 \pi)^3} \int dP_1 dP_2 dP_3 (-g)^{-1/2} \frac{P_\mu P_\nu}{P^0} f(x^i, P_j,\tau).
\end{equation}
Thus, in terms of the new momentum variable ${\mathbf q}$, we find for the components of the stress--energy tensor to first order in the metric perturbations
\begin{eqnarray}
T^0_{\ 0} & = &  - a^{-4} \int \frac{d^3 q}{(2 \pi)^3} \epsilon (q,\tau) f({\mathbf x},{\mathbf q},\tau), \nonumber \\
T^0_{\ i} & = & a^{-4} \int \frac{d^3 q}{(2 \pi)^3} q f({\mathbf x},{\mathbf q},\tau)=  - T^i_{\ 0}, \nonumber \\
T^i_{\ j} & = &  a^{-4} \int \frac{d^3 q}{(2 \pi)^3} \frac{q^i q_j}{\epsilon}
f({\mathbf x},{\mathbf q},\tau),
\end{eqnarray}
where $\epsilon(q,\tau)  \equiv a E = \sqrt{a^2 m^2 + q^2}$, and $E$ is the energy measured by a stationary observer with respect to the coordinates~(\ref{eq:conformal}).

The  phase space density $f({\mathbf x},{\mathbf q},\tau)$ can be tracked with  the Boltzmann equation,
\begin{equation}
\label{eq:boltzmann}
\frac{\partial f}{\partial \tau} + \frac{d {\mathbf x} }{d \tau} \cdot \frac{\partial f}{\partial {\mathbf x}} + 
\frac{d {\mathbf q}}{d \tau} \cdot \frac{\partial f}{\partial {\mathbf q}} = C[f],
\end{equation}
where the term $C[f]$ encapsulates the effects of non-gravitational interactions.  Because we are  interested in the evolution of perturbations in the neutrino fluid after they decouple from the cosmic plasma, this term can be neglected.  
Equation~(\ref{eq:boltzmann})  is nonlinear.    We therefore resort to perturbation theory, and
expand the phase space density as $f({\mathbf x},{\mathbf q},\tau) = f_0(q) + f_1({\mathbf x},{\mathbf q},\tau) + \cdots$,
where $f_0(q)$ is the phase space density of the fluid if it were in a homogeneous and isotropic FLRW background.  For a neutrino fluid, the homogeneous and 
isotropic component of its phase density is
$f_0(q) = [1+\exp(q/T_{\nu,0})]^{-1}$, where $T_{\nu,0}$ is the neutrino temperature today, (cf.\  equation~(\ref{eq:relfd})).  Then, expanding out the Boltzmann equation~(\ref{eq:boltzmann}), we 
find for the first order correction,
\begin{equation}
\frac{\partial f_1}{\partial \tau} + \frac{\mathbf q}{\epsilon} \cdot \frac{\partial f_1}{\partial {\mathbf x}} + 
\frac{d q}{d \tau}
\frac{\partial f_0}{\partial q} =0,
\end{equation}
or, equivalently in Fourier space,
\begin{equation}
\label{eq:boltzmann2}
\frac{\partial  \tilde{f}_1}{\partial \tau} + i\frac{q}{\epsilon} ({\mathbf k} \cdot \hat{\mathbf q} ) \tilde{f}_1 + 
[q \dot{\tilde\phi} - i \epsilon ({\mathbf k} \cdot \hat{\mathbf q}) \tilde\psi ]
\frac{\partial f_0}{\partial q} =0,
\end{equation}
where $\hat{\mathbf q} \equiv {\mathbf q}/q$, and the expression for $dq/d\tau$ arises from solving for the metric~(\ref{eq:conformal})
the geodesic equation,
$d P^\alpha/d \tau = - (1/m) \Gamma^{\alpha}_{\beta \gamma} P^{\beta} P^\gamma$.

Consider first the case of CDM. Generally we are more  interested in bulk quantities 
 than in the detailed phase space distribution.   In the case of CDM, two essential bulk quantities are the density contrast and the velocity divergence,
\begin{equation}
\tilde\delta_c = \frac{1}{\bar{n}_c} \int \frac{d^3q}{(2 \pi a )^3} \tilde{f}_1,\qquad
\tilde\theta_c  =  \frac{1}{\bar{n}_c} \int \frac{d^3q}{(2 \pi a)^3} \tilde{f}_1 \frac{q}{\epsilon} ({\mathbf k} \cdot \hat{\mathbf q}),
\end{equation}
where $\bar{n} \equiv (2 \pi a )^{-3} \int d^3 q f_0$ is the mean CDM number density.
Evidently, $\tilde\delta_c$ and $\tilde\theta_c$  are simply kinetic moments of the phase space distribution.  We can therefore construct 
equations of motion for these quantities directly by taking kinetic moments of the Boltzmann equation~(\ref{eq:boltzmann2}) itself.  
The resulting equations for the zeroth and first moments are
 \begin{equation}
 \dot{\tilde\delta}_c + \tilde\theta_c -3 \dot{\tilde\phi}=0,  \qquad \dot{\tilde\theta}_c + {\cal H} \tilde\theta_c - k^2 \tilde\psi = 0,
 \end{equation}
 which are none other than  the linearised continuity and Euler equations respectively.  Note that in order to write the Euler equation in this form, we have had to neglect those phase space integrals containing powers of $(q/\epsilon)^n$ 
 with $n>1$.  This is a justified approximation for CDM since CDM is by definition nonrelativistic.   By the same argument, all higher kinetic moments are negligible.

For neutrinos, these approximations are clearly ill-justified, especially at early times when they are relativistic.  Even at late times it is important that we keep track of kinetic moments beyond the first, because it is precisely these higher moments that give rise to the non-clustering effects!
These considerations inevitably lead us to the conclusion that we must solve the full Boltzmann equation~(\ref{eq:boltzmann2}) for a neutrino fluid.  This can be done by brute force integration of the equation  for a large sample in momentum space.  A cleverer route, however, is to recast equation~(\ref{eq:boltzmann2}) into a so-called Boltzmann hierarchy.

The Fourier space phase space density  $\tilde{f}_1$ is nominally a function of the wavevector ${\mathbf k}$, momentum ${\mathbf q}$, and conformal time $\tau$.  However, an inspection of equation~(\ref{eq:boltzmann2}) reveals that aside from the dependence on the magnitudes $k$ and $q$, $\tilde{f}_1$ 
has no explicit dependence on the direction of either vector except through the scalar 
product $\mu \equiv \hat{\mathbf k} \cdot \hat{\mathbf q}$.  Thus without loss of generality we can write $\tilde{f}_1 = \tilde{f_1} (k,q,\mu,\tau)$, and decompose $\tilde{f}_1$ in terms of  a Legendre series,
\begin{eqnarray}
\label{eq:moments}
&&\tilde{f}_1(k,\mu,q,\tau) = \sum_{n=0}^\infty (-i)^n (2 n+1) \Psi_n(k,q,\tau) P_n(\mu), \nonumber \\
&&\Psi_n(k,q,\tau) = \frac{1}{2 (-i)^n} \int_{-1}^1 d\mu \  \tilde{f}_1(k,\mu,q,\tau) P_n(\mu),
\end{eqnarray}
with $P_n(\mu)$  the Legendre polynomial of degree $n$.

The advantage of this decomposition is manifest once we recognise that each moment $\Psi_n$ can in fact be related to a bulk quantity of the neutrino fluid.  
From the zeroth moment we recover the energy density and pressure perturbations,
\begin{equation}
\tilde{\delta \rho}_\nu = a^{-4} \int \frac{d^3 q}{(2 \pi)^3} \epsilon \Psi_0, \qquad
\tilde{\delta p}_\nu =  \frac{1}{3} a^{-4} \int \frac{d^3 q}{(2 \pi)^3} \epsilon \left(\frac{q}{\epsilon}\right)^2 \Psi_0,
\end{equation}
while the velocity divergence and the anisotropic stress,
\begin{eqnarray}
(\bar{\rho}_\nu + \bar{p}_\nu)\tilde\theta_\nu &=&  k a^{-4} \int \frac{d^3 q}{(2 \pi)^3} \epsilon \left(\frac{q}{\epsilon}\right) \Psi_1, \nonumber \\
(\bar{\rho}_\nu + \bar{p}_\nu) \tilde\sigma_\nu &=&  \frac{2}{3} a^{-4} \int \frac{d^3 q}{(2 \pi)^3} \epsilon \left(\frac{q}{\epsilon}\right)^2 \Psi_2,
\end{eqnarray}
are linked to the first and second moments respectively.
A set of evolution equations for the moments $\Psi_n$ can now be constructed by integrating  equation~(\ref{eq:boltzmann2}) as per equation~(\ref{eq:moments}):
\begin{eqnarray}
\dot{\Psi}_0 &=& -\frac{qk}{\epsilon} \Psi_1 - \dot{\phi} \frac{d f_0}{d \ln q}, \nonumber \\
\dot{\Psi}_1 & = & \frac{qk}{3 \epsilon} (\Psi_0 - 2 \Psi_2) - \frac{\epsilon k}{3 q} \psi \frac{d f}{d \ln q}, \nonumber \\
\dot{\Psi}_{n \geq 2}& =& \frac{qk}{(2 n + 1) \epsilon} [ n \Psi_{n-1} - (n+1) \Psi_{n+1}].
\end{eqnarray}
The is the so-called Boltzmann hierarchy.
The multipole at which to truncate the hierarchy
depends on the desired level of accuracy.   In the  Boltzmann code CAMB~\cite{Lewis:1999bs},
1000  momentum bins and $n_{\rm max} = 6$ appear to suffice for the computation of the matter power 
spectrum accurate to the percent level.

\subsection{Initial conditions}

Initial conditions for the perturbations are set by inflation (see, e.g., \cite{riotto}).  
In the inflationary paradigm, a scalar field---the inflaton---whose dynamics is dominated by its potential energy
drives an early phase of exponential expansion.  During this expansion, quantum fluctuations on the scalar field are stretched to superhorizon scales ($k \ll {\cal H}$) and imprinted on the spacetime metric. At the end inflation, the inflaton releases its kinetic energy through particle production, a process known as reheating.  When the kinetic energy of these particles come to dominate the energy budget, 
the universe enters into the radiation-domination era.

Because the particles thus produced originate from the inflaton field, they must also inherit its superhorizon fluctuations.  In the simplest scenario, the single-field inflation scenario, in which only one scalar field is responsible for both driving the exponential expansion and seeding the metric perturbations, the generic predictions for the initial 
perturbations are~\cite{mabertschinger}
\begin{eqnarray}
\label{eq:adiabatic}
&& \delta_\gamma= \delta_\nu = \frac{4}{3} \delta_c = \frac{4}{3} \delta_b = - 2 \psi, \qquad \theta_\gamma = \theta_\nu = \theta_c = \theta_b = \frac{1}{2} k^2 \tau \psi, \nonumber \\
&&\sigma_\nu = \frac{1}{15} (k \tau)^2 \psi, \qquad
\psi = \frac{20 C}{15 + 4 R_\nu}, \qquad \phi = \left(1 +\frac{2}{5} R_\nu \right) \psi,
\end{eqnarray}
where $C$ is some constant determined by the amplitude of the fluctuations from inflation, and $R_\nu = \bar{\rho}_\nu/(\bar{\rho}_\nu + \bar{\rho}_\gamma)$ during radiation domination.  These are called adiabatic perturbations, and can be interpreted to mean that the number density ratios between the various particle species are the same everywhere.    For the neutrino Boltzmann hierarchy, equation~(\ref{eq:adiabatic}) implies that
$\Psi_0 = -(\tilde\delta_\nu/4)  (d f_0/d \ln q)$, $\Psi_1  =  - (\tilde\theta_\nu/3 qk)   (d f_0/d \ln q)$,
and $\Psi_2 =  -( \tilde\sigma_\nu/2)  (d f_0/d \ln q)$.

Inflation does not predict exactly what initial values $\tilde\delta(k)$, $\tilde\psi(k)$,and so on should adopt.  Rather, it 
predicts that these quantities are random in nature, and their statistics are quantified by the $n$-point correlators.  Of particular 
importance is the two-point correlator, e.g., $\langle \tilde \delta({\mathbf k}) \tilde\delta({\mathbf q}) \rangle = \delta_D({\mathbf k}+{\mathbf q}) P_\delta (k)$,
where $P_\delta(k)$ is called the power spectrum of $\tilde\delta$.
Single-field inflation models generically predict $P(k)$
to be the only nontrivial statistics: all odd correlators are vanishingly small, while all even correlators can be constructed from $P(k)$.
These are known as gaussian initial conditions.   Note that $P(k)$ depends only on the magnitude and not the direction of ${\mathbf k}$ 
because of statistical homogeneity and isotropy.

\section{Neutrino mass and cosmological observables\label{sec:present}}

\subsection{Observational probes and signatures of neutrino mass}

Massive neutrinos impact on cosmological observables such as the CMB and the LSS in two distinct ways.
Firstly, neutrino masses in the sub-eV to eV range significantly alter the expansion history near the 
epoch of matter--radiation equality.  Secondly, neutrino free-streaming affects the
growth of structures at late times.  In this section, I describe the manifestation of 
these effects in the CMB and the LSS as well as various observational probes currently used to 
detect these signatures of neutrino mass.

\subsubsection{Cosmic microwave background anisotropies}

The state-of-the-art instrument used 
to measure the CMB temperature and polarisation anisotropies
is the Wilkinson Microwave Anisotropy Probe (WMAP), which
measures the full-sky anisotropies with a resolution of $0.5$ degrees~\cite{Jarosik:2010iu}.
Anisotropies on smaller scales  have been measured by an array of 
ground-based and balloon experiments such as ACBAR~\cite{Reichardt:2008ay}, BICEP~\cite{Chiang:2009xsa},
QuAD~\cite{Brown:2009uy}, and ACT~\cite{Das:2010ga}.

\begin{figure}[t]
\begin{center}
  \includegraphics[width=13cm]{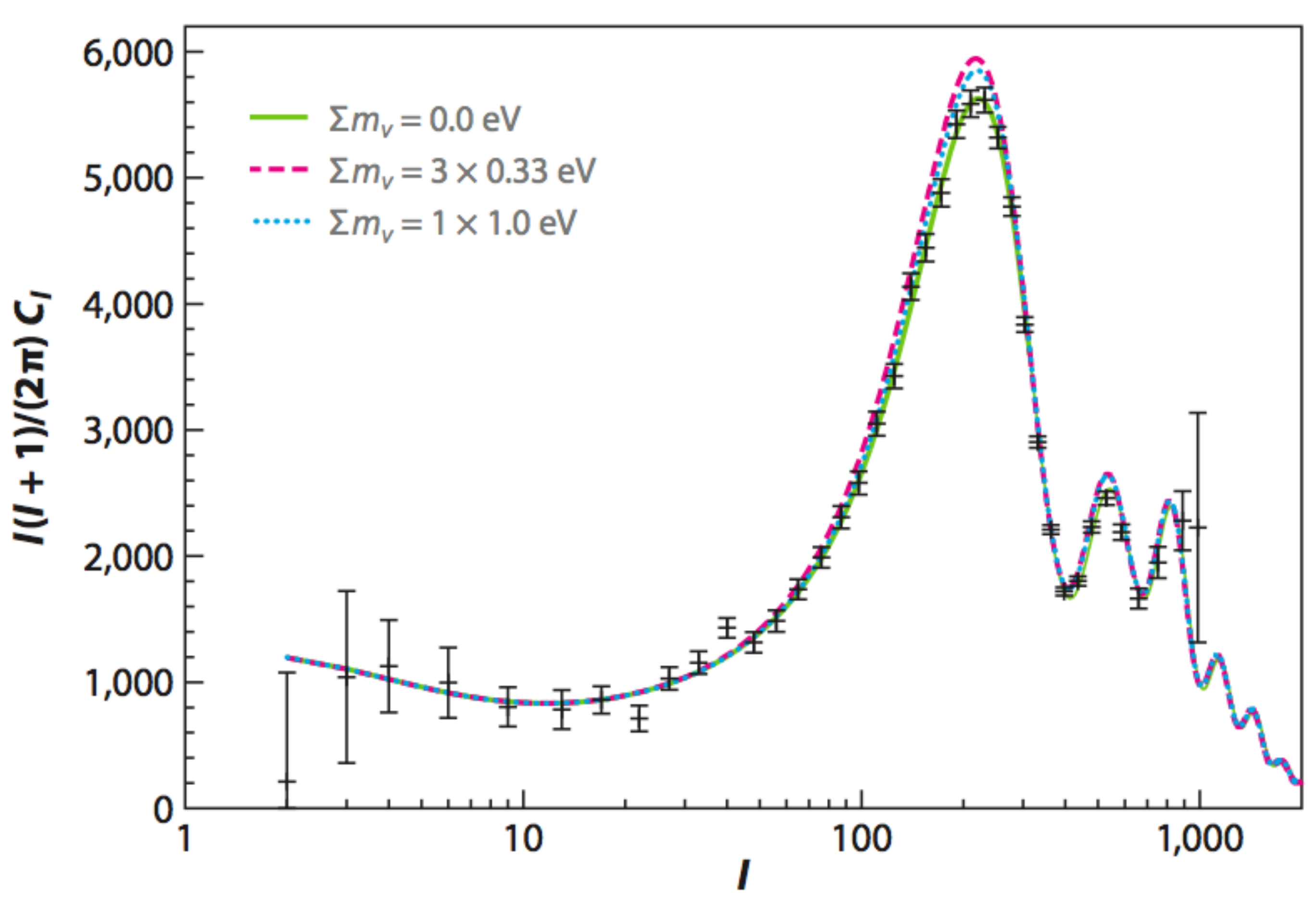}
  \end{center}
  \caption{The cosmic microwave background temperature anisotropy spectrum for a model with massless neutrinos and two models with massive neutrinos.
Data are from WMAP after three years of observation~\cite{Hinshaw:2006ia}. \label{fig:cmb}}
\end{figure}

Because these fluctuations are mapped onto the surface of a sphere, it
is convenient to decompose them in terms of spherical harmonics $Y_{\ell m}$ (cf.\ Fourier decomposition in flat space).
For example, for the temperature fluctuations,
\begin{eqnarray}
&& \frac{\Delta T_\gamma}{T_\gamma} (\theta,\phi) = \sum_{\ell=0}^{\infty} \sum_{m=-\ell}^{\ell} a^T_{\ell m} Y_{\ell m} (\theta,\phi),
\nonumber \\
&&a^T_{\ell m} = (-i)^\ell 4 \pi \int d \Omega \ Y_{\ell m}^*(\theta,\phi) \frac{\Delta T_\gamma}{T_\gamma} (\theta,\phi),
\end{eqnarray}
where $\theta$ and $\phi$ denote the angular position of a measurement, and $d \Omega$  the solid angle.
The statistics of the fluctuations are quantified at the lowest order by a  two-point correlator
$\langle a^X_{\ell m} a^{Y*}_{\ell'm'} \rangle = \delta_{\ell \ell'} \delta_{m,m'} C^{XY}_\ell$,
where $X,Y=T,E,B$ denotes temperature, and the $E$ and $B$ mode polarisation.%
\footnote{Polarisation of the CMB is  decomposed into a curl-free $E$-mode and a divergence-free $B$-mode.  The 
latter is present at linear order only if there are primordial gravitational waves (tensor modes) or vector modes, and has yet to be detected.}
Note that $C^{XY}_\ell$ has no dependence on the index $m$ because of statistical homogeneity and isotropy.
 Figure~\ref{fig:cmb} shows the temperature auto-correlation $C^{TT}_\ell$
for several different cosmologies.

 The main signature of neutrino masses in the CMB  comes about via the so-called early integrated 
Sachs--Wolfe (ISW) effect.  Photon decoupling occurs at a temperature of $T \sim 0.26$~eV.  
Shortly before this event, at $T \sim 1$~eV, is the epoch of matter--radiation
equality.  The transition from radiation to matter 
domination induces a nontrivial evolution for the metric perturbations $\phi$ and $\psi$, and hence  the photon geodesics.
Because this transition occurs so close in time to the epoch of photon decoupling, remnants of this nontrivial evolution 
are preserved in the CMB anisotropies, especially in the height of the first acoustic peak of 
the CMB temperature power spectrum.  This is the early ISW effect.

If the absolute neutrino mass  is roughly of order $1$~eV, then the relic neutrinos also transit 
from being a fully relativistic particle species to a nonrelativistic one at $T \sim 1$~eV, thereby contributing 
to the early ISW effect. Figure~\ref{fig:cmb} shows examples of the CMB temperature
power spectrum for cosmological models with massive neutrinos, juxtaposed with the prediction of a
 model with massless neutrinos.  This enhancement of the first peak enables us to put a constraint on the absolute neutrino mass
scale from the CMB alone.

\subsubsection{Probes of large-scale structure}  

The idea of ``weighing neutrinos'' with LSS probes 
hinges on two effects~\cite{het}. The first effect is the manifestation in the matter power spectrum 
of the background effect described above associated with the 
 transition from radiation to matter domination.  This effect is 
most important for sub-eV to eV mass.  The second effect is 
free-streaming.  Neutrinos are born with a large
thermal velocity, which makes them difficult to capture into potential wells corresponding 
to wavenumbers larger than  the so-called free-streaming wavenumber~\cite{Ringwald:2004np},
$k_{\rm fs} = 1.5 \sqrt{\Omega_m h^2} (1+z)^{-1/2} (m_\nu/{\rm eV}) \  {\rm Mpc}^{-1}$,
even after they have become nonrelativistic.
In other words, if we were to replace part of the CDM content with neutrinos, then 
free-streaming would cause the growth of the overall density perturbations at $k > k_{\rm fs}$ to be less efficient.

\begin{figure}[t]
\begin{center}
  \includegraphics[width=13cm]{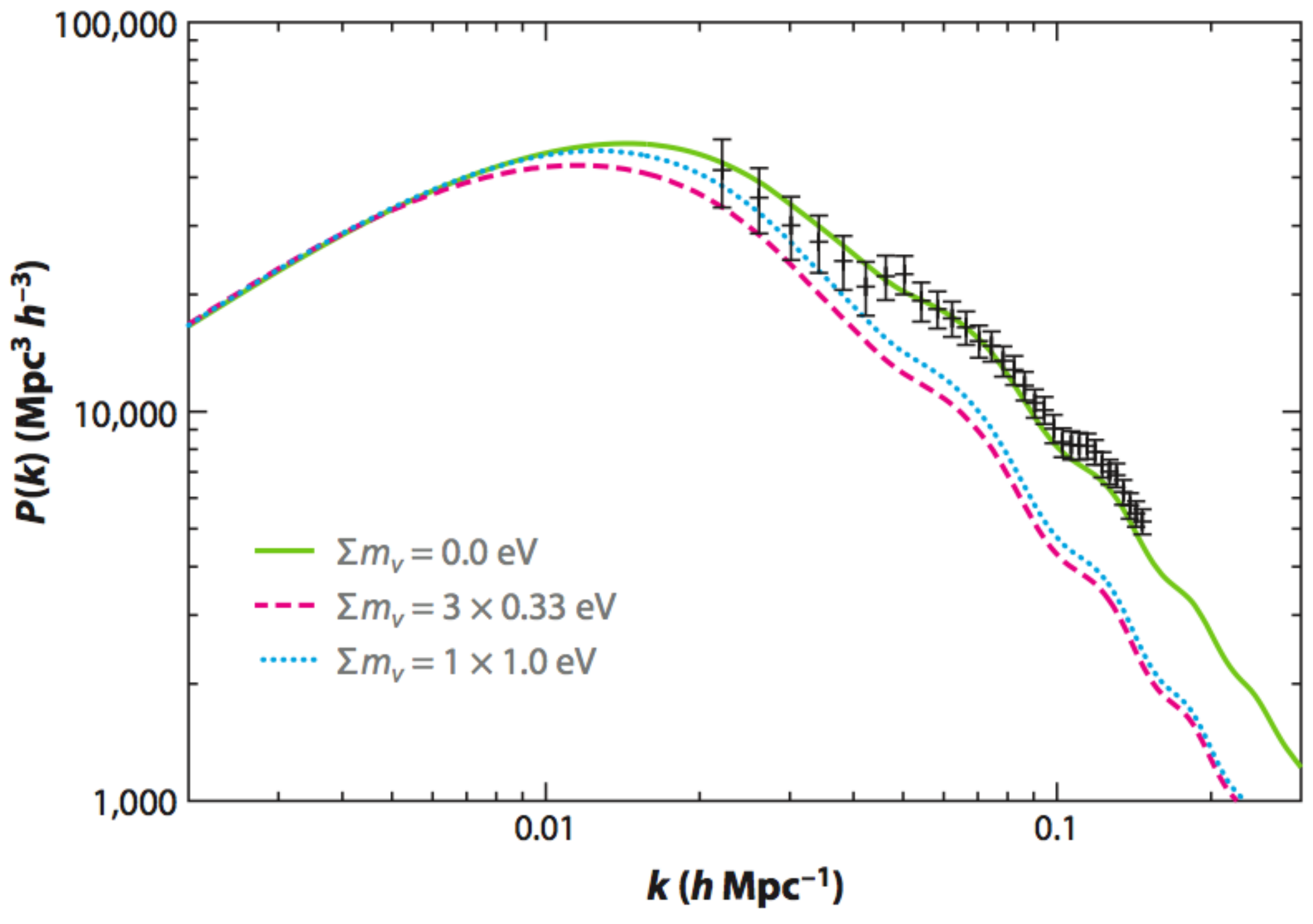}
 \end{center} 
  \caption{The large-scale matter power spectrum for a model with massless neutrinos and two models with massive neutrinos.
The data points
  are from the 2dF survey \cite{Cole:2005sx}.\label{fig:lss}}
\end{figure}

In the large-scale matter power spectrum, these two effects combine to produce a 
 suppression of power on small scales scaling roughly as
 $\Delta P/P  \sim 8 f_\nu$,
where $f_\nu \equiv \Omega_\nu/\Omega_m$
is the neutrino fraction,
according to linear perturbation theory.
  A detailed derivation can be found in reference~\cite{Lesgourgues:2006nd}.
Figure~\ref{fig:lss} shows this suppression for the case of $\sum m_\nu=1 \ {\rm eV}$.  Note that the suppression 
relative to the massless neutrino case depends primarily on the sum $\sum m_\nu$.  Exactly how the neutrino masses are distributed 
contributes only a minor effect.

Several types of surveys of the large-scale matter distribution are available, each employing different tracers of the matter density field.
Currently, galaxy redshift surveys are  the most powerful LSS probe, with the 
the Sloan Digital Sky Survey (SDSS) \cite{Abazajian:2008wr}  being the largest to date.
 These surveys produce essentially galaxy catalogues, from which
 one can compute the galaxy number density at each point in space, and hence their $n$-point statistics.  The present 
 generation of surveys observe at  relatively low redshifts ($z<0.5$) and at 
small length scales ($k > 0.02 \ h \ {\rm Mpc}^{-1})$ just beyond the reach of CMB observations.

Because galaxy surveys measure the galaxy number density fluctuations, in order to use these measurements to constrain 
cosmology, we must assume that the clustering of galaxies traces that of the underlying matter distribution 
 up to a constant bias factor, i.e., $\delta_{\rm gal} = b
\delta_{\rm m}(k)$.   The value of  $b$ depends on the galaxy sample used, and is usually marginalised at the end of an
 analysis because it cannot be computed from first principle.
This procedure works well on large  scales, 
but is expected to fail on small scales.   Indeed, analysis of the luminous red galaxy (LRG) sample of the SDSS 
already requires that we introduce some form of nonlinear modelling to account for  the so-called scale-dependent
bias at $k> 0.1 \ h \ {\rm Mpc}^{-1}$~\cite{tegmarklrg,Reid:2009xm,bias}.  However, because the nuisance parameters of the bias model are always  
marginalised at the end of the analysis, 
this procedure is effectively the same as discarding all (nonlinear) data beyond  $k> 0.1 \ h \ {\rm Mpc}^{-1}$~\cite{bias,Hamann:2010pw}.

Another LSS probe that has gained considerable attention recently is the Lyman-$\alpha$ (Ly$\alpha$) forest.  In essence these are 
absorption features in intergalactic low density gases along the line of sight to a distant quasar.
Since density fluctuations in these gases are expected to follow the underlying matter density field, 
measurements of the Ly$\alpha$ flux power spectrum can be used to
reconstruct the matter power spectrum on small scales at large redshifts ($z> 1$).
However, unlike the case of the galaxy bias, the gas--matter bias is highly nontrivial, and shows
strong dependence on the gas temperature and ionisation history,
 metal line contamination, and so on.  So far 
 various working groups have not reached a consensus regarding 
 the implications of
Ly$\alpha$ for precision cosmology~\cite{McDonald:2004xn,Viel:2005eg,Viel:2005ha,Viel:2006yh}.

A number of other probes, such as weak gravitational lensing and cluster abundance, have also been explored to constrain cosmology.
However, since these probes tend to rely strongly on nonlinear physics which is presently not very well understood, I defer their
 discussion to section~{\ref{sec:future}.

\subsubsection{Other probes}

Standard candles and standard rulers are objects of known luminosity and known physical size respectively.
Measurement of a standard candle's apparent magnitude allows us to deduce its luminosity distance. 
Similarly, the angular size of a standard ruler reveals its angular diameter distance.
If the redshifts of a series of standard candles or rulers are known, then it is possible to establish the
 distance--redshift relation, and hence  the low-redshift expansion history of the universe.  
The most well-known standard candles to date are the Type Ia supernovae (SN), while the position of 
the baryon acoustic oscillations (BAO) peak is  a standard ruler frequently used in precision cosmology~\cite{Percival:2009xn}.

Another important measurement is the present expansion rate of the universe, 
quantified by the present day Hubble parameter $H_0$.  Its value---at least that in our immediate neighbourhood ($z<0.1$)---has been established from
observations with the Hubble Space Telescope (HST) to be 
$H_0 = 74.2 \pm 3.6 \ {\rm km} \ {\rm s}^{-1} \  {\rm Mpc}^{-1}$~\cite{Riess:2009pu}.

Neutrino masses have little effect on the distance relations and the universal expansion at low redshifts.  Therefore these probes  play no direct role 
in constraining the absolute neutrino mass scale.  Nonetheless, they remain useful for  removing 
parameter degeneracies.

\subsection{Cosmological model}

The simplest model required to account for all cosmological 
observations to date is the concordance flat $\Lambda$CDM model. 
 It makes the following assumptions.
 \begin{enumerate}
\item General relativity holds on all length scales.
\item  The large-scale spatial geometry of the universe is flat.
\item The energy content of the universe comprises photons, whose 
 energy density is fixed by the COBE FIRAS measurement of the CMB temperature and energy spectrum~\cite{Fixsen:1996nj},
$3.046$ families of thermalised massless neutrinos with temperature given by 
 equation~(\ref{eq:tnu}), and
unspecified amounts of baryons, CDM, and vacuum energy due to a cosmological constant.
\item The initial conditions, i.e., the statistics and amplitude of the primordial scalar perturbations to the 
FLRW metric, are set by the simplest, single-field inflation models.  The perturbations are gaussian and adiabatic, and the presence of primordial 
 vector or tensor modes is negligible.
The power spectrum of density perturbations is described by a constant spectral index $n_s$ and an amplitude $A_s$, i.e., $P_\delta (k) = A_s k^{n_s}$, 
both of which are free to vary.

\end{enumerate}

In addition to  these parameters, a minimal model of the CMB anisotropies requires an additional astrophysical parameter, the optical depth to reionisation $\tau$.  
In all there are six free physical parameters in the ``vanilla'' model of cosmology:
\begin{equation}
\Omega_b h^2, \Omega_{\rm CDM} h^2, \Omega_\Lambda, n_s, A_s,\tau.
\end{equation}
Spatial flatness implies $\Omega_b + \Omega_{\rm CDM} + \Omega_\Lambda=1$, from which 
we can deduce the reduced Hubble parameter $h$.  

Of course, these six parameters are not the only ones that can affect cosmology: there are many more.
These other parameters are not included in ``standard'' analyses because they do not change 
significantly the goodness-of-fit, i.e., the best fit $\chi^2$ of the model does not improve by more than 3 or
4 with the inclusion of any one extra parameter.  This improvement is deemed too modest by many practitioners to 
justify the inclusion of extra parameters, although it is unclear where one should draw the line.
Indeed, some of these extra parameters, such as the neutrino mass, are  physically extremely well motivated.

The simplest neutrino mass limits can be obtained by including in the vanilla description a variable amount of neutrino dark matter characterized by 
the sum of the neutrino masses $\sum m_\nu$.   We call this the ``vanilla+$m_\nu$'' model.

\subsection{Present constraints and model dependence}

Using the effects discussed above, many authors have placed bounds on the sum of the neutrino masses.
I discuss some of them below.

\subsubsection{Vanilla bounds}

Within the  vanilla+$m_\nu$ framework,  the 7-year data from WMAP  (WMAP7) already limit the sum of the neutrino masses to
$\sum m_\nu < 1.3$~eV (95\% credible interval)~\cite{Komatsu:2010fb}.   
Including small-scale CMB experiments in the analysis leads 
 to virtually no improvement~\cite{Hannestad:2010yi}.
Incorporating distance information from BAO~\cite{Percival:2009xn} tightens the bound to $\sum m_\nu < 0.85$~eV~\cite{Hannestad:2010yi}, 
while combining CMB with the SDSS LRG halo power spectrum (HPS)~\cite{Reid:2009xm} gives $\sum m_\nu < 0.61$~eV~\cite{Hannestad:2010yi}.%
\footnote{Despite their sensitivity to neutrino free-streaming,
LSS measurements cannot constrain a 7-parameter model on their own, and
must be used together  with  large-scale CMB observations.}

Using also the HST measurement of $H_0$ further strengthens the bound.  For example, CMB+HST yields $\sum m_\nu < 0.58$~eV (95\% C.I.)~\cite{Hannestad:2010yi}, 
while combining CMB+HPS+HST gives $\sum m_\nu < 0.44$~eV~\cite{Komatsu:2010fb,Hannestad:2010yi,Reid:2009nq}. 
Further improvement can be expected from the inclusion of SN data.  Reference~\cite{Thomas:2009ae}
finds  $\sum m_\nu < 0.28$~eV (95\% C.I.)
for WMAP5+SN+BAO+MegaZ\-+HST, where MegaZ is a SDSS photometric redshift catalogue  and SN refers to the first year 
data set of SNLS~\cite{astier}.  However, one should exercise caution here regarding the use of SN data,
for  they appear at present to be plagued by unresolved systematic issues associated with the light-curve fitting methods~\cite{sn};
Discrepant light curve fitters have led to estimates  of dark energy parameters that differ  by as much as  $2.5 \sigma$.
 
Several other LSS probes have also produced interesting constraints on $\sum m_\nu$,  although they are arguably more strongly dependent on nonlinear physics than those quoted above.
For example, reference~\cite{Ichiki:2008ye}
finds $\sum m_\nu < 1.1$~eV (95\% C.I.) from weak lensing observations with
the CFHTLS together with WMAP5.    Another analysis combines the cluster mass function from weak  lensing with WMAP3 to obtain
$\sum m_\nu < 1.43$~eV (95\% C.I)~\cite{kristiansen}.  A  recent variant uses the X-ray cluster abundance from the ROSAT All Sky Survey and 
the Chandra X-ray observatory with WMAP5+SN+BAO to find
$\sum m_\nu <  0.33$~eV (95\% C.I.)~\cite{Mantz:2009rj}.

\subsubsection{Extended models}

Cosmological models beyond vanilla+$m_\nu$ fall into two categories: those that ``perturb'' around vanilla+$m_\nu$, and those 
that represent a complete change of paradigm.

\paragraph{Perturbations around vanilla}

Models that perturb around vanilla+$m_\nu$
include the introduction of (i) extended inflation physics, (ii) extra relativistic species,  (iii) dynamical dark energy/modified gravity,
(iv) non-flat spatial geometry.  Each type of model is discussed below.

(i) Popular extensions to the vanilla description of the primordial 
perturbations include a running spectral index (i.e., scale-dependent $n_s$), the presence of primordial gravitational waves,
and isocurvature modes (from, e.g., multi-field inflation).  The latter two affect only the CMB
anisotropies at low multipoles and are not directly degenerate with neutrino masses~\cite{Zunckel:2006mt}.
A running spectral index may  mimic or offset to an extent the small-scale suppression in the matter power spectrum 
caused by neutrino free-streaming.  However, any such fine-tuning necessarily leads to modifications of the large-scale CMB
anisotropies and can therefore be constrained.  Indeed, constraints on $\sum m_\nu$ from WMAP5 alone are already completely independent of these additional features~\cite{Reid:2009nq}.

(ii) Solutions to several outstanding particle physics puzzles predict the production of additional light particles in the early universe.
Amongst these are light sterile neutrinos and hot dark matter axions.  Such particles contribute relativistic energy density to the total energy budget, leading to 
$N_{\rm eff}>3.046$.  As a CMB fit parameter, $N_{\rm eff}$ is highly degenerate with $H_0$, $\Omega_m h^2$ and indirectly $\sum m_\nu$,
 so that in order to obtain constraints on $\sum m_\nu$ it is necessary to introduce additional data sets.  The best available bound for the vanilla+$m_\nu$+$N_{\rm eff}$ model
 is $\sum m_\nu <  0.89$~eV (95\% C.I.), coming from CMB+HST+BAO~\cite{Hamann:2010pw}.  The corresponding preferred value of $N_{\rm eff}$ is $4.47^{+1.82}_{-1.74}$.

(iii) Dynamical dark energy scenarios involve replacing the cosmological constant with 
a fluid whose equation of state parameter $w_{\rm DE}$ satisfies  $w_{\rm DE}< -1/3$ and may additionally be time-dependent~\cite{copeland}.
More elaborate variants usually involve coupling between the dark energy and other matter components (e.g., \cite{LaVacca:2009yp}).
Phenomenologically,  $w_{\rm DE}$  exhibits considerable degeneracy 
with the neutrino mass~\cite{Hannestad:2005gj}.  For a constant $w_{\rm DE}$, the best bound comes from WMAP7+HPS+HST, 
$\sum m_\nu < 0.71$~eV (95\% C.I.) for vanilla+$m_\nu$+$w_{\rm DE}$~\cite{Komatsu:2010fb} (note that the CMB alone cannot constrain scenarios with a free $w_{\rm DE}$).  This is to be compared with 
$\sum m_\nu < 0.44$~eV for $w_{\rm DE}=-1$ using the same data sets.  If in addition $N_{\rm eff}$ is allowed to vary, then the bound relaxes to 
$\sum m_\nu < 1.16$~eV (95\% C.I.)~\cite{Hamann:2010pw}.

Scenarios that  modify  general relativity at very 
large distances are primarily constructed to explain the observed 
late-time accelerated expansion of the universe in lieu of a cosmological constant.
 Phenomenologically
they share many similarities with dynamical dark energy scenarios.

An interesting issue concerns the possibility that certain coupled dark energy and modified gravity scenarios 
can lead to scale-dependent clustering and thus mimic 
neutrino free-streaming~\cite{starobinsky}.  This issue has yet to be explored in detail.
 
(iv)  Flat spatial geometry is one of the pillars of the inflationary paradigm; Relaxing the assumption of spatial flatness is perhaps the least theoretically
well-motivated modifications to the vanilla model.  Phenomenologically, however, a nonzero spatial curvature does appear to be degenerate with $\sum m_\nu$~\cite{GonzalezGarcia:2010un}, in the same way that $w$ is degenerate with $\sum m_\nu$.

\paragraph{Different paradigm}
Several cosmological models that differ radically from concordance $\Lambda$CDM have been proposed 
in the literature, including broken scale invariance~\cite{Blanchard:2003du} 
and inhomogeneous void models that seek to explain current 
cosmological observations without invoking a phase of late-time accelerated expansion. 
Although these models have found some success with subsets of the available data, it is generally difficult to
reconcile the simplest variants with all data sets.  The role of massive neutrinos has yet to be elucidated for 
inhomogeneous models.  The model of~\cite{Blanchard:2003du}, although incompatible with BAO and SN,
requires  three species of neutrinos, each which has a mass of $\sim 0.5$~eV~\cite{Hunt:2007dn}.
This value an be tested by the tritium $\beta$-decay
 experiment KATRIN~\cite{katrin}.

\section{Nonlinear regime I: $N$-body simulations\label{sec:nbody}}

Linear perturbation theory  has  served  the current generation of observational probes well.
To advance further, however, future LSS probes will begin to observe on smaller length scales, where
we expect the evolution of  density perturbations to enter  the mildly nonlinear regime.  At the same time, 
scales that have  already been 
observed will be probed with ever increasing precision. 
Thus in order to maximise the potential of future probes to constrain cosmology, we must go beyond linear order
for accurate computations of the observables from theory. The length scales of interest here fall into the range $k \sim 0.1 \to 1 \ {\rm Mpc}^{-1}$.

To this end, two lines of approach, $N$-body simulations and semi-analytic techniques, 
are currently being nvestigated in an attempt to bring the large-scale matter power 
spectrum under control at the percent level.  The former is arguably the more
definitive approach of the two and its validity generally extends to much smaller length scales than are possible with the latter.  However, 
where they are valid, semi-analytic techniques do in general offer a significant reduction in computation time, and are in many ways 
more transparent.  The two approaches therefore complement each other.

In this section I describe the basic idea of how to compute nonlinear corrections to the LSS distribution, especially in the presence
of massive neutrinos, by way of $N$-body simulations.  For a review of $N$-body methods, see, e.g., reference~\cite{Bertschinger:1998tv}.
Semi-analytic techniques will be discussed in section~\ref{sec:analytic}.
Note that most current nonlinear investigations neglect the effects of baryonic physics, i.e.,
baryons are treated as a collisionless matter component like CDM, while scattering processes
responsible for the heating and cooling of gasses, star formation, and so on are ignored.    The  common conception is that these effects  are 
important only for  $k > 1 \ {\rm Mpc}^{-1}$.   
Since we are concerned mainly with the mildly nonlinear scales $k \sim 0.1 \to 1 \ {\rm Mpc}^{-1}$,
it seems justified to ignore these effects.   However, although baryonic physics is anticipated to be a minor player on the scales of interest, it is not inconceivable that some effect is present, even if only at the percent level~\cite{Jing:2005gm}.   
Therefore, one should hear in mind that although the nonlinear techniques to be discussed below are all valid and 
correct in their own right, they probably represent only part of the full picture of nonlinear corrections.

\subsection{Basic idea of an $N$-body simulation}

$N$-body simulations are arguably the most definitive approach to nonlinear clustering to date.
In this section, I described the basic building blocks of an $N$-body simulation.

\subsubsection{Cosmological simulations}

Modern cosmological simulations are performed using a  cubic ``box''  typically with a comoving side length of $\sim10 0 \to 1000$~Mpc.  This box is filled with a 
fluid (or several fluids) whose phase space density $f_\alpha$ evolves according to the Boltzmann equation~(\ref{eq:boltzmann}) with the collision term $C[f_\alpha]$ set to zero.

At $z< 100$, all particle species of interest are nonrelativistic.   Relativistic energy densities and the associated anisotropic stresses are negligibly small, so that
the metric perturbations $\psi$ and $\phi$ are  equal.
Furthermore, we expect nonlinear evolution to be important only on subhorizon scales ($k \gg {\cal H}$). 
These considerations justify the use of a pseudo-Newtonian description for the system, in the sense that apart from the Hubble expansion, all other aspects of gravitational physics are to be described by Newton's laws.  In practice this means
\begin{equation}
\label{eq:characteristics}
\frac{d {\mathbf x}}{d \tau} = \frac{\mathbf q}{m_\alpha a}, \qquad \frac{d {\mathbf q}}{d \tau} = - a m_\alpha \nabla \phi,
\end{equation}
so that the collisionless Boltzmann equation can now be written as
\begin{equation}
\label{eq:vlasov}
\frac{\partial f_\alpha}{\partial \tau} + \frac{{\mathbf q}}{m_\alpha a} \cdot 
\nabla f_{\alpha}   - a m_\alpha \nabla \phi \cdot 
\frac{\partial f_{\alpha}}{\partial {\mathbf q}}=0,
\end{equation}
where the Poisson equation
\begin{equation}
\label{eq:poisson}
\nabla^2 \phi = 4 \pi G a^2 \sum_\alpha \bar{\rho}_\alpha \delta_\alpha = \frac{3}{2}{\cal H}^2 \sum_\alpha \Omega_\alpha (\tau) \delta_\alpha
\end{equation}
links the density perturbations to the gravitational potential, 
with $\Omega_\alpha(\tau) \equiv \bar{\rho}_\alpha (\tau)/\rho_{\rm crit}(\tau)$.

The collisionless Boltzmann equation expresses conservation of phase space density along characteristics $\{ {\mathbf x}(\tau),{\mathbf q}(\tau)\}$.
An fluid element drawn initially from $\{ {\mathbf x}_i,{\mathbf q}_i\} \to \{ {\mathbf x}_i+ d{\mathbf x},{\mathbf q}_i+ d {\mathbf q}\} $
will move according to the characteristic equations~(\ref{eq:characteristics}).
Thus tracking  characteristics originating from all possible  $\{ {\mathbf x}_i,{\mathbf q}_i\}$ is
equivalent to solving the collisionless Boltzmann equation.  
In practice one can only track a small  sample of characteristics.  This is then an $N$-body simulation.

The Boltzmann equation~(\ref{eq:vlasov}) implies that fluid elements should be sampled from both coordinate and momentum space.
However, since the emergence of the concordance $\Lambda$CDM model as the standard model of  cosmology, a default 
cosmological simulation of CDM requires only that we sample from coordinate space.
This is a a reasonable approximation;  Already at linear order we have seen that the CDM fluid is 
treated as a single stream fluid described only by its density contrast and bulk velocity.  The velocity dispersion is by definition zero.

More precisely, we can take kinetic moments of the phase space density,
\begin{eqnarray}
\rho_\alpha  &\equiv&  m_\alpha \int \frac{d^3 q}{(2 \pi a )^3} f_\alpha, \quad {\rm  (0th\  moment)} \nonumber \\
\rho_\alpha {\mathbf u}  &\equiv&  m_\alpha \int \frac{d^3 q}{(2 \pi a )^3} \frac{\mathbf q}{m_\alpha a}  f_\alpha, \quad {\rm  (1st\  moment)}\nonumber \\
\rho_\alpha \sigma_{\alpha, ij} &\equiv &  m_\alpha  \int \frac{d^3 q}{(2 \pi a )^3} \frac{q_i q_j}{(m_\alpha a)^2}  f_\alpha - \rho_\alpha u_i u_j, \quad {\rm  (2nd\  moment)}
\end{eqnarray}
and of the Boltzmann equation~(\ref{eq:vlasov}),
\begin{eqnarray}
\label{eq:fluid}
&&\dot{\delta}_\alpha + \nabla \cdot [(1+\delta_\alpha) {\mathbf u}_\alpha] = 0, \quad {\rm  (0th\  moment)} \nonumber \\
 &&\dot{\mathbf u}_\alpha + {\cal H} {\mathbf u}_\alpha + ({\mathbf u}_\alpha \cdot \nabla) {\mathbf u}_\alpha + \nabla \phi + \frac{1}{\rho_\alpha} \nabla_j( \rho_\alpha \sigma_{\alpha, ij})=0. \quad {\rm  (1st\  moment)} 
 \end{eqnarray}
 Simulating the CDM case simply consists of setting the stress term $\sigma_{c,ij}$ to zero, thereby truncating the hierarchy at the first moment equation.  The remaining zeroth and first moments equations are  none but the continuity and Euler equation in Eulerian fluid dynamics.  Defining the convective derivative (i.e., the time derivative taken along a path moving with velocity ${\mathbf u}_\alpha$),
$D/D \tau \equiv \partial/\partial \tau + {\mathbf u}_\alpha \cdot \nabla$,
we see that the Euler equation can be written as
$D {\mathbf u}_\alpha /D \tau+ {\cal H} {\mathbf u}_\alpha = - \nabla \phi$.
 This equation is exactly equivalent to the characteristic equations~(\ref{eq:characteristics}) if we identify ${\mathbf u}_\alpha = d{\mathbf x}/d \tau$.
In other words, an element of CDM fluid moving with velocity ${\mathbf u}$ evolves according to Newton's second law of motion.

Thus a  recipe for simulating  CDM  emerges. Imagine a cubic  box of comoving  volume $V$ filled with a fluid whose initial mass density is $\bar{\rho}$ .  This box is divided into $N$ blocks, with each block represented by a point particle of mass $m = \bar{\rho}/NV$.  These particles move under 
one another's gravity according to equation~(\ref{eq:characteristics}), where Hubble expansion has been absorbed into the comoving coordinates.   Periodic boundary conditions are imposed, meaning that one should envisage the box as one amongst an infinite lattice of identical boxes; Should a trajectory lead outside the box on one side, it would reenter from the opposite side.

\subsubsection{Initial conditions}

Simulations are usually started at high redshifts, typically  $z=50 \to 100$, where perturbation theory is still valid, and 
  can be used to set the initial conditions of a simulation.
The most popular  implementation is by way of the Zel'dovich approximation and its variants.

Imagine a simulation box spanned by a 3D grid, where we place on each grid point point particles of identical masses. 
The particles thus positioned represent a uniform fluid.   To factor in the initial density contrasts and bulk velocities, we assign to each particle a random initial displacement ${\mathbf s}({\mathbf r})$ from the grid point ${\mathbf r}$ and an initial velocity vector ${\mathbf v}$ as  dictated by the (perturbative) clustering statistics at the time the simulation is initiated $\tau_i$.  In the Zel'dovich approximation, the displacement vector ${\mathbf s}$ is given by
$\nabla \cdot {\mathbf s} = - \hat{\delta}({\mathbf r},\tau_i)/D^+(\tau_i)$,
where $D^+(\tau_i)$ is the linear growth function at the initial time, $\hat{\delta}({\mathbf r},\tau_i)$ is a random number drawn from a gaussian distribution with variance 
determined by the linear power spectrum  $P(k,\tau_i)$,  and $\nabla \times {\mathbf s} =0$.
Thus, the initial position of the simulation particle and its velocity are now given by
\begin{equation}
{\mathbf x} = {\mathbf r} + D^+(\tau_i) {\mathbf s} ({\mathbf r}), \qquad
{\mathbf v}= \frac{d D^+}{d \tau} {\mathbf s} ({\mathbf r}).
\end{equation}
The randomly displaced particles now represent an inhomogeneous fluid whose clustering statistics respects the predictions of linear perturbation theory;

The Zel'dovich approximation is technically the first order solution to the fluid equations~(\ref{eq:fluid}) in Lagrangian perturbation theory.
In modern $N$-body simulations, the displacement vectors and initial velocities are generally computed from second order Lagrangian perturbation theory.

\subsubsection{Gravity solver}

The heart of an $N$-body code is the gravity solver, which computes  the gravitational force felt by each simulation particle.
Direct summation of the  forces between each pair of particles may be intuitive.  
However, because the number of operations scales as $N^2$ where $N$ is the number of  particles, 
this approach quickly gets out of hand in cosmological simulations in which $N$ often exceeds a million.

Two broad classes of gravity solvers are employed in modern simulations: the Particle--Mesh (PM) code, and the Tree code.  The PM code is particularly important for cosmological simulations  because of the ease with which one can implement periodic boundary conditions.

\paragraph{Particle--Mesh code}

The PM code uses a grid or mesh on which to evaluate the gravitational forces.  The presence of a mesh enables the use of discrete Fourier transforms via the Fast Fourier Transform algorithm, a pivotal routine in this class of gravity solver.   Here, for the purpose of evaluating $\phi({\mathbf x})$,  the particles are envisaged as clouds and then smoothed on to the grid points.  The total mass density $\rho({\mathbf x})$ on each grid point is then a sum of each particle's contribution.  This density is then Fourier transformed into 
$\tilde{\rho}({\mathbf k})$, which allows for an algebraic evaluation of the gravitational force (in Fourier space) ${\cal F}[\nabla \phi] $ via the Poisson equation~(\ref{eq:poisson}), i.e., 
${\cal F}[\nabla \phi] = i {\mathbf k} \tilde{\phi} = -  i {\mathbf k} 4 \pi G a^2 \tilde{\rho}/k^2$.
This value is inverse Fourier transformed back to real space to give $\nabla \phi$ on each grid point, and then interpolated to the particle positions to be fed into the characteristic equations~(\ref{eq:characteristics}).  The total number of operations scales as $N_m \log N_m$, where $N_m$ is the number of grid points,  
significantly less than that required in a direct summation scheme.  The  disadvantage, however, is that the force resolution becomes unreliable on 
length scales close to and below the grid spacing.

\paragraph{Tree code}

To illustrate the principle of the Tree code, take as an example three particles A, B and C, where  B and C are separated by a distance $s$, and 
 $d$ denotes the distance from A to the centre-of-mass (c.o.m.) of B and C.  To evaluate the combined force of B and C on A, the Tree algorithm stipulates that 
 we treat B and C as one particle centred on their c.o.m.\ if their angular separation satisfies $s/d < \theta$, where $\theta$ is some small number of order unity.  
 If $s/d > \theta$, then B and C must be treated as two separate particles.  This idea of lumping two particles together according to their angular separation can be 
 easily generalised to the multi-particle case, and has the advantage that  the total number of operations scales as $N \log N$.  
 A popular implementation is the Barnes--Hut algorithm~\cite{Barnes:1986nb}. 
  However, periodic boundary conditions  are not intrinsic to the Tree algorithm
  as they are to the PM code.
   Nonetheless, the Tree algorithm is often used in combination with 
  the PM code in cosmological simulations 
 to overcome the small-scale shortcomings  of the latter, such as in the publicly available $N$-body code GADGET-II~\cite{gadget}.

\subsection{$N$-body simulations with massive neutrinos}

Simulation with massive neutrinos have many parallels with pure CDM simulations, but are complicated by the fact
that neutrinos always come with large thermal velocities.  In this section, I describe how large neutrino thermal
 velocities translate into complications for $N$-body simulations and review some recent efforts towards meeting 
 these challenges.

\subsubsection{Thermal neutrino motion}

Simulating  massive neutrinos is nontrivial because of their large thermal velocities.  Recall that the unperturbed neutrino 
phase space density is described by a relativistic Fermi--Dirac distribution. For a $1$~eV neutrino, equation~(\ref{eq:thermal}) shows that
the average thermal velocity is easily 15\% the speed of light at $z=50$ when we start a simulation, even though the bulk velocity of the neutrino fluid  
 is much less than the Hubble flow.

To simulate (nonrelativistic) massive neutrinos, we employ the same equations of motion as those for simulating a CDM fluid.  Relativistic corrections are generally too small for the desired level of accuracy to be of concern.  The only difference between neutrinos and CDM arises when  
the initial conditions are implemented: in addition to sampling from coordinate space, it is  necessary to 
sample from momentum space for a neutrino fluid to account for its 
velocity dispersion.  A simple implementation introduced in reference~\cite{Brandbyge:2008rv}  uses the Zel'dovich approximation to generate an initial random bulk velocity for the neutrino fluid 
at each grid point.  Added to this bulk velocity is a thermal component consisting of a random thermal speed drawn from the relativistic Fermi--Dirac distribution 
and a random unit vector specifying the direction of the thermal contribution.

 Figure~\ref{fig:damping} shows the large-scale matter power spectrum for several massive neutrino cosmologies from the simulations of reference~\cite{Brandbyge:2008rv}, 
 including the aforementioned neutrino thermal velocity dispersion (see also reference~\cite{Viel:2010bn}).  More precisely, this figure  shows the suppression  of power induced 
 in the matter power spectrum when we replace part of the CDM content with massive neutrinos.  
 Importantly, although  linear perturbation theory predicts a relative suppression of $\Delta P/P \sim 8 f_\nu$ in the
  matter power spectrum on small scales due to neutrino free-streaming, 
the $N$-body results in Figure~\ref{fig:damping} indicate that the suppression in fact scales 
more like $\Delta P/P \sim 10 f_\nu$!

 \begin{figure}[t]
 \begin{center}
  \includegraphics[width=13cm]{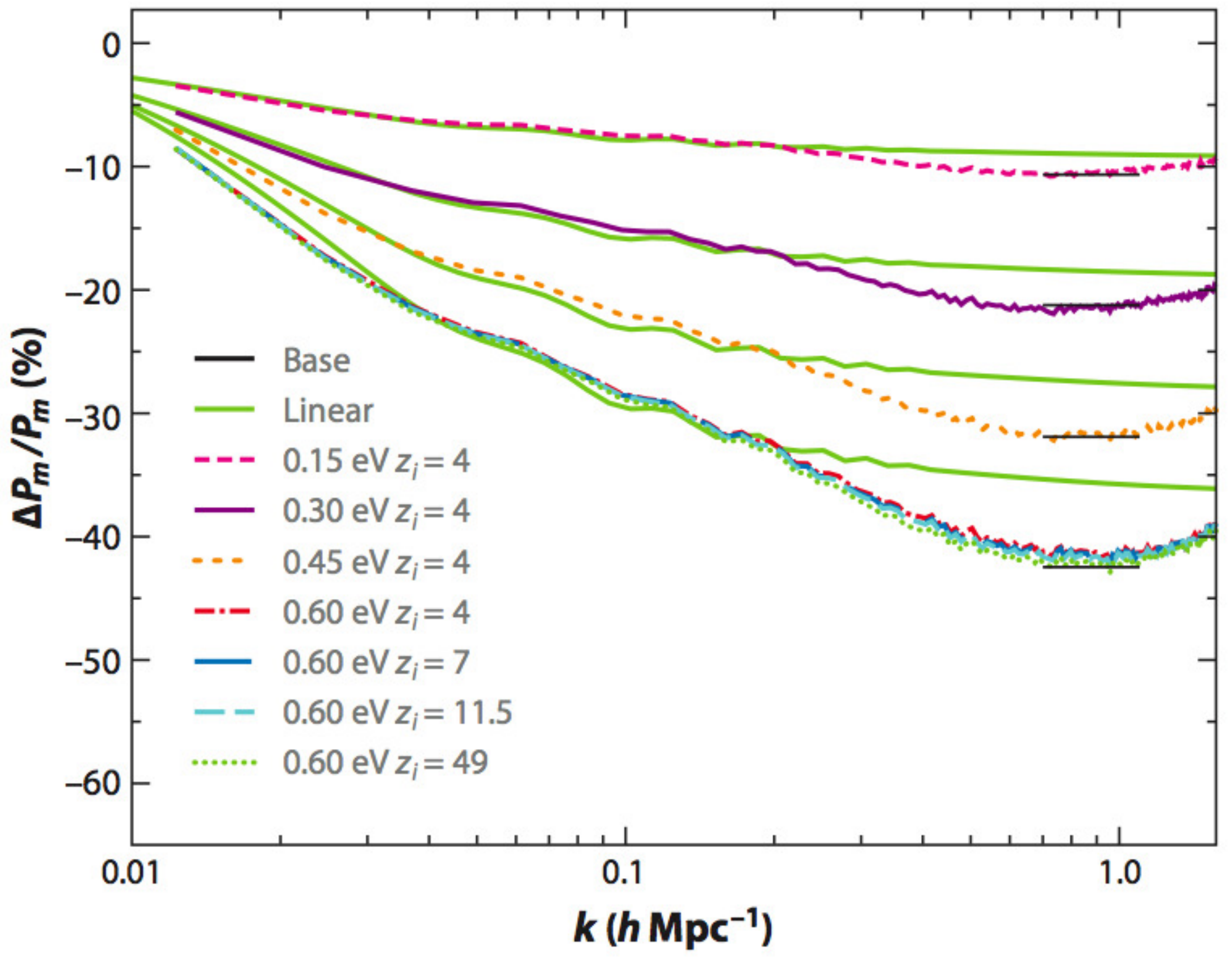}
 \end{center} 
  \caption{The large-scale matter power spectrum for massive neutrino cosmologies relative to the case
with massless neutrinos from $N$-body simulations.  
The solid/green lines represent the predictions from linear perturbation theory, while the colourful dotted/dashed
lines are the results from full-scale nonlinear simulations for different neutrino masses ($\sum m_\nu$)  indicated on the plot.  
Figure reproduced from reference~\cite{Brandbyge:2008rv}.~\label{fig:damping}}
\end{figure}

\subsubsection{Approximate schemes}

Because of the additional random sampling involved, simulations including massive neutrinos can be plagued by an increased level of shot noise unless the sampling  is adequate.  Also, the neutrinos' large thermal velocities  mean that the time steps taken by the simulation code must be appropriately small in order to accurately track the  trajectories.    Starting the simulation at a lower redshift  when the neutrinos have slowed down somewhat 
alleviates these problems to a good extent
but at the expense of reduced  accuracy.%
\footnote{A high initial redshift is required if one is to compute the absolute matter power spectrum to percent accuracy. 
 However, if one is interested merely in the 
relative suppression due to neutrino free-streaming, then starting a simulation as late as $z=4$ does not affect the results~\cite{Brandbyge:2008rv}.}
All in all,  simulating massive neutrinos can be very computationally intensive and time-consuming, so
 approximate schemes are worth consideration.

\paragraph{Grid-based linear neutrino perturbations}

As the name implies,  in this scheme the neutrino perturbations are assumed to remain at linear order and  evolved on each grid point in a PM 
code using a linear Boltzmann code such as CAMB~\cite{Lewis:1999bs}, while the CDM fluid is given a particle realisation.    The neutrinos' contribution to the total mass density is then
 added to the CDM's contribution on each grid point, so that the gravitational force felt by each CDM  particle can be evaluated via the usual PM machinery.  
This scheme was explored in reference~\cite{Brandbyge:2008js} and found to be accurate to the percent level
for $\sum m_\nu < 1$~eV compared with a full-fledged simulation.   However, the reduction in the simulation run time can be as much as a factor of ten.
Note that in reference~\cite{Brandbyge:2008js} nonlinear corrections to the CDM perturbations are not fed back into the evolution equations 
of the linear neutrino perturbations, which may be  a point to improve upon.

\paragraph{Hybrid scheme}

A more adventurous hybrid scheme was introduced in reference~\cite{Brandbyge:2009ce}.  
Here, the neutrino perturbations for all momenta are initially tracked with linear perturbation theory,
 a reasonable approximation given that at high redshifts the neutrino thermal speeds are so large that neutrino clustering must be minimal.  As the  neutrinos slow down at low 
 redshifts, beginning with the low momentum states, 
 those neutrinos with a momentum-to-temperature ratio below
some predefined threshold are converted into  $N$-body particles upon crossing.%
\footnote{In practice, owing to the Eulerian nature of Boltzmann codes like CAMB and the Lagrangian nature of $N$-body simulations,     
a bin by bin conversion is a nontrivial accounting problem.  A simple solution that works well is to convert a large number of momentum bins at the same time.
See discussion in reference~\cite{Brandbyge:2009ce}.}

This scheme avoids the initial shot noise problem, but  at the same time allows us to track the neutrino nonlinearities fully at late times down to very small length scales.
It is thus particularly well-suited to the study of such small scale problems such as halo density profiles and the halo mass function.  The latter counts the number density 
of virialised structure (``halos'') per unit mass at a given redshift, and was computed for massive neutrino cosmologies for the first time via $N$-body simulations in
reference~\cite{Brandbyge:2010ge}.  Figure~\ref{fig:cluster} shows the halo mass function for several cosmologies.

\begin{figure}[t]
\center
\includegraphics[width=13cm]{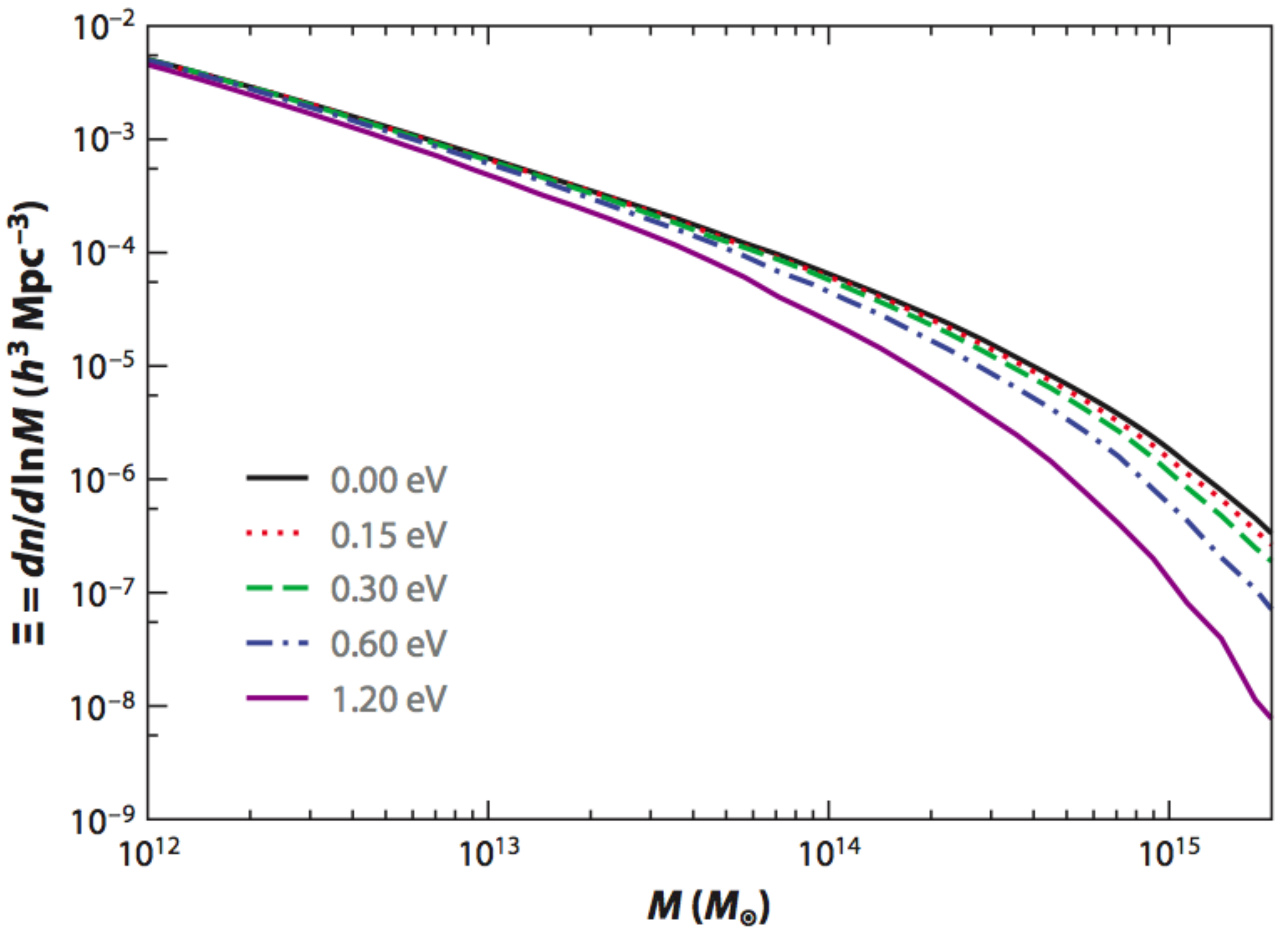}
\center
\caption{Halo mass function for cosmologies with the indicated neutrino masses ($\sum m_\nu$). The 0.0~eV case corresponds
to the $\Lambda$CDM model.  Figure reproduced from reference~\cite{Brandbyge:2010ge}.\label{fig:cluster}}
\end{figure}

\section{Nonlinear regime II: Semi-analytic methods\label{sec:analytic}}

Numerical simulations are arguably the most definitive way to date to treat the nonlinear clustering problem. 
 However, 
time, they can also be very computationally demanding and time consuming, especially for cosmologies beyond the $\Lambda$CDM model.
For this reason, a number of semi-analytic schemes have been investigated for the calculation of nonlinear corrections to the matter power spectrum
and other LSS observables.  I describe two below.

\subsection{Higher Order Perturbation Theory}

As the name implies, the higher order perturbation theory approach consists of going beyond linear when solving 
the equations that describe the evolution of the density perturbations.

\subsubsection{The equations}

The starting equations in this approach are the fluid equations~(\ref{eq:fluid}) and the Poisson equation~(\ref{eq:poisson}) 
already introduced earlier (for a detailed review of perturbative methods, see, e.g., \cite{Bernardeau:2001qr}).
Because the main constituent of the matter content is CDM, it makes sense to consider first nonlinearities in the CDM 
fluid, while keeping the neutrino perturbations 
at  linear order.   
This is the approach taken in references~\cite{Saito:2008bp,Wong:2008ws}, and is a 
reasonable approximation supported by $N$-body simulation results~\cite{Brandbyge:2008js}  for neutrino masses not exceeding $\sum m_\nu \sim 1 $~eV.
 There have also been attempts to model the nonlinearities in the neutrino fluid~\cite{Shoji:2009gg}.

Cold dark matter is pressureless at linear order so that  $\sigma_{c, ij}^{(1)}=0$.  At very nonlinear scales, however,  
we do expect pressure and anisotropic stress to be generated.  
However, because we are interested only in the mildly nonlinear regime ($k \sim 0.1 \to  1 \  {\rm Mpc}^{-1}$),  these 
factors are probably  not very important.    Therefore the first approximation is to assume that $\sigma_{c,ij}=0$ 
at all orders.  It follows that in the absence of a finite $\sigma_{c,ij}$, no vorticity ${\mathbf w}_c \equiv \nabla \times {\mathbf u}_c$ can be generated for the CDM fluid.

With these assumptions, we can now rewrite the continuity and the Euler equations in terms of the only remaining degrees of freedom, the density contrast $\delta_c$ and 
the velocity divergence $\theta_c  \equiv \nabla \cdot {\mathbf u}_c$.   Since we are working  with perturbation theory, it is convenient to rewrite these equations in Fourier space,
\begin{eqnarray}
\label{eq:fouriereq}
&&\dot{\tilde\delta}_c({\mathbf k},\tau) +\tilde\theta_c ({\mathbf k},\tau) = - \int d^3 q_1 d^3 q_2 \delta_D({\mathbf k} - {\mathbf q}_{12})
\alpha ({\mathbf q}_1,{\mathbf q}_2) \tilde\theta_c({\mathbf q}_1,\tau) \tilde\delta_c ({\mathbf q}_2,\tau),  \nonumber\\
&&  \dot{\tilde\theta}_c({\mathbf k},\tau) \! +\! {\cal H} \tilde\theta_c({\mathbf k},\tau)  -k^2 \tilde\phi ({\mathbf k},\tau) = \nonumber \\
&& \hspace{25mm} \int  d^3 q_1 d^3 q_2  \delta_D({\mathbf k}- {\mathbf q}_{12})\beta ({\mathbf q}_1,{\mathbf q}_2) \tilde\theta_c({\mathbf q}_1,\tau) \tilde\theta_c ({\mathbf q}_2,\tau),
\end{eqnarray}
where ${\mathbf q}_{12} = {\mathbf q}_1 + {\mathbf q}_2$, and the vertices $\alpha({\mathbf q}_1,{\mathbf q}_2)  = {\mathbf q}_{12} \cdot {\mathbf q}_1/q_1^2$ and
$\beta ({\mathbf q}_1,{\mathbf q}_2) = q_{12}^2 ({\mathbf q}_1 \cdot {\mathbf q}_2)/(2 q_1^2 q_2^2)$
encode the coupling between Fourier modes.

At this point, we define a doublet
$\varphi \equiv ( 
		    \tilde\delta_c, \;
		    -\tilde\theta_c /{\cal H} )^T$,
and a new time variable $s \equiv \ln a$.
Then the equations of motion~(\ref{eq:fouriereq}) can be recast into  a  compact form,
\begin{equation}
\label{eq:compact}
\partial_s \varphi_a({\mathbf k}, s)+ \Pi_{ab}({\mathbf k}, s) \varphi_b ({\mathbf k}, s)= \int  d^3 q_1 d^3 q_2 \gamma_{abc} ({\mathbf k},{\mathbf q}_1,{\mathbf q}_2)\varphi_b({\mathbf q}_1,s) \varphi_c ({\mathbf q}_2,s),
\end{equation}
where
\begin{equation}
\Pi ({\mathbf k}, s)= \left[ \begin{array}{cc}
	  0 & -1 \\
	  - \frac{3}{2} [ \Omega_c(s) + \sum_{\alpha \neq c} \Omega_\alpha (s) \frac{\tilde\delta_\alpha ({\mathbf k}, s)}{\tilde\delta_c ({\mathbf k}, s)}]
	  & 1+\frac{1}{s} \frac{d {\cal H}}{d s} \end{array} \right],
\end{equation}
and the vertex function $\gamma_{abc}$ is zero except for
\begin{eqnarray}
\gamma_{121}  ({\mathbf k},{\mathbf q}_1,{\mathbf q}_2)&=&\gamma_{112}  ({\mathbf k},{\mathbf q}_2,{\mathbf q}_1)= \delta_D({\mathbf k} - {\mathbf q}_{12}) \frac{1}{2} \alpha ({\mathbf q}_1,{\mathbf q}_2), \nonumber \\
 \gamma_{222} ({\mathbf k},{\mathbf q}_1,{\mathbf q}_2) &=& \delta_D({\mathbf k} - {\mathbf q}_{12})\beta ({\mathbf q}_1,{\mathbf q}_2). 
\end{eqnarray}
Let us call equation~(\ref{eq:compact}) the ``master'' equation, as we will return to it below.

In standard perturbation theory, we solve  equation~(\ref{eq:compact})  using a  perturbative series,
$\varphi({\mathbf k},s) = \sum_{n=1}^{\infty} \varphi^{(n)} ({\mathbf k},s)$.
The equation of motion for the $n$-th order perturbation is therefore
\begin{eqnarray}
&&\partial_\tau \varphi^{(n)}_a({\mathbf k},s)+ \Pi_{ab} ({\mathbf k},s)\varphi^{(n)}_b ({\mathbf k},s)=\nonumber \\
&& \hspace{10mm}  \int  d^3 q_1 d^3 q_2 \gamma_{abc} ({\mathbf k},{\mathbf q}_1,{\mathbf q}_2)
\sum_{m=1}^{n-1} \varphi^{(n-m)}_b({\mathbf q}_1,s) \varphi^{(m)}_c ({\mathbf q}_2,s),
\end{eqnarray}
which has the formal solution
\begin{eqnarray}
\label{eq:formal}
\varphi^{(n)}_a({\mathbf k},s) \!\!\!\!&= & \!\! \!\! g_{ab} ({\mathbf k}, s,s_i) \varphi^{(n)}_b({\mathbf k},s_i) \\
&+&\!\!\!\! \int \! d^3 q_1 \!  \int \! d^3 q_2 \! 
\int_{s_i}^s \!  d s'  g_{ab} ({\mathbf k}, s,s') \gamma_{bcd} \!
\sum_{m=1}^{n-1} \varphi^{(n-m)}_c({\mathbf q}_1,s') \varphi^{(m)}_d ({\mathbf q}_2,s').\nonumber 
\end{eqnarray}
Following field theory language, $g_{ab}({\mathbf k}, s, s')$ is known as the linear propagator.  For a matter-dominated universe, i.e., $\Omega_c(s)=1$, 
$g_{ab}=g_{ab}(s, s')$ is independent of $k$, and takes the form
\begin{equation}
g_{ab}(s,s') = \frac{e^{s-s'}}{5} \left( \begin{array}{cc}
			  3 & 2 \\
			  3 & 2 \end{array} \right)
-\frac{e^{-3(s-s')/2}}{5} \left( \begin{array}{cc}
			  -2 & 2 \\
			  3 & -3 \end{array} \right).
\end{equation}
The first term is a growing solution; The astute reader will have recognised that this 
 is but the linear growth function for a matter-dominated universe.  The second term is a decaying solution.

The formal solution~(\ref{eq:formal}) in its present form has little practical use.  To make it more  user-friendly, we introduce some simplifications:
(i) We assume that initially all higher order terms are vanishingly small; only $\varphi^{(1)}({\mathbf k},s_i)$ is nonzero;
(ii) Linear solutions are always in the growing mode.  In other words, whenever we encounter a linear term
$\varphi^{(1)}_a({\mathbf k},s)$, we make the approximation
$\varphi^{(1)}_a({\mathbf k},s) = e^{s-s_i} \varphi^{(1)}_a({\mathbf k},s_i)$; and
(iii) The initial time of the time integral is in the infinite past, i.e., $s_i = -\infty$.
 Then,  an order by order evaluation shows that
\begin{equation}
\label{eq:solution}
\varphi^{(n)}_a ({\mathbf k},s)= \int d^3 q_1 \cdots d^3 q_2 \delta_D({\mathbf k} - {\mathbf q}_{1\cdots n}) 
Q^{(n)}_a{\mathbf q}_1, \ldots, {\mathbf q}_n) \varphi^{(1)}_1 ({\mathbf q}_1,s)\cdots \varphi^{(1)}_1 ({\mathbf q}_n,s),
\end{equation}
where the kernels are given by
\begin{eqnarray}
\label{eq:kernels}
Q^{(n)}_a{\mathbf q}_1, \ldots, {\mathbf q}_n)  
&=&\sigma^{(n)}_{ab} \sum_{m=1}^{n-1} \gamma_{bcd}({\mathbf k},{\mathbf q}_{1\cdots m},{\mathbf q}_{m+1\cdots n}) \nonumber \\
&& \times Q^{(m)}_c({\mathbf q}_1, \ldots, {\mathbf q}_m)
Q^{(n-m)}_c({\mathbf q}_{m+1}, \ldots, {\mathbf q}_n),
\end{eqnarray}
with
\begin{equation}
\sigma_{ab}^{(n)} =
\frac{1}{(2n+3) (n-1)} \left[ \begin{array}{cc}
				2n+1 & 2 \\
				3 & 2n \end{array} \right].
\end{equation}
Note that in older literature, the kernels are denoted as $F_{n} \equiv Q^{(n)}_1$ and
$G_{n} \equiv Q^{(n)}_2$.

\subsubsection{The power spectrum}

We are interested in the clustering statistics of matter as quantified by the matter power spectrum
$\langle \tilde\delta_m({\mathbf k}, \tau) \tilde\delta_m({\mathbf q}, \tau) \rangle =  \delta_D({\mathbf k}+{\mathbf q}) P_m (k,\tau)$,
where the matter density contrast $\tilde\delta_m ({\mathbf k}, \tau) = (1-f_\nu) \tilde\delta_c({\mathbf k}, \tau) +f_\nu \tilde\delta ({\mathbf k}, \tau)$
counts both contributions from CDM and neutrinos.  Thus $P_m(k,\tau)$ is a linear sum of the CDM and the neutrino auto-correlation spectra
$P_c(k,\tau)$ and $P_\nu(k,\tau)$, 
and their cross-correlation spectrum $P_{c\nu}(k,\tau)$,
\begin{equation}
P_m (k,\tau) =  (1-f_\nu)^2 P_c(k,\tau) + 2  (1-f_\nu) f_\nu P_{c \nu} (k,\tau)+ f_\nu^2 P_\nu(k,\tau).
\end{equation}
Because we are keeping the neutrino perturbations at  linear order, we  need only to consider nonlinear corrections to $P_c(k,\tau)$ and $P_{c \nu}(k,\tau)$.

As with the density perturbations,
 the power spectrum can also be expressed as an infinite sum.  For the CDM auto-correlation, for example, we find
\begin{equation}
 P_c (k,\tau) =  P_{c}^{11} (k,\tau) +  P_c^{22} (k,\tau) + P_c^{13} (k,\tau) \cdots,
\end{equation}
with
\begin{eqnarray}
\delta_D({\mathbf k}+{\mathbf q}) P_c^{11} (k,\tau)&=&\langle \tilde\delta^{(1)}_c({\mathbf k}, \tau) \tilde\delta^{(1)}_c({\mathbf q}, \tau) \rangle,\nonumber \\
\delta_D({\mathbf k}+{\mathbf q}) P_c^{22} (k,\tau)&=&\langle \tilde\delta^{(2)}_c({\mathbf k}, \tau) \tilde\delta^{(2)}_c({\mathbf q}, \tau) \rangle,\nonumber \\
\delta_D({\mathbf k}+{\mathbf q}) P_c^{13} (k,\tau)&=&\langle \tilde\delta^{(1)}_c({\mathbf k}, \tau) \tilde\delta^{(3)}_c({\mathbf q}, \tau) \rangle 
+ \langle \tilde\delta_c^{(3)}({\mathbf k}, \tau) \tilde\delta_c^{(1)}({\mathbf q}, \tau) \rangle.
\end{eqnarray}
Here, the leading order term $P^{11}_c(k,\tau)$ is the usual linear power spectrum, while $P^{22}_c+P^{13}_c$ are collectively known as the one-loop correction.
Similarly for the CDM--neutrino cross-correlation,
\begin{equation}
 P_{c\nu} (k,\tau) =  P_{c\nu}^{11} (k,\tau) + P_{c\nu}^{13} (k,\tau),
\end{equation}
with 
$\delta_D({\mathbf k}+{\mathbf q}) P_{c\nu}^{13} (k,\tau)=\langle \tilde\delta^{(3)}_c({\mathbf k}, \tau) \tilde\delta^{(1)}_\nu({\mathbf q}, \tau) \rangle$.
The ``22'' term is absent here because the neutrino perturbations are assumed to be linear.

Given the solutions~(\ref{eq:solution}),  we can readily evaluate the higher order terms:
\begin{eqnarray}
\label{eq:powerspectra}
P_c^{22} (k) &=& 2 \int d^3 {\mathbf q} \ [F_2^{(s)}({\mathbf k}-{\mathbf q},{\mathbf q})]^2 P_c^{11} (|{\mathbf k}-{\mathbf q}|)  P_c^{11}(q), \nonumber \\
P_c^{13} (k) &=& 6 \int d^3 {\mathbf q} \ F_3^{(s)}({\mathbf k},{\mathbf q},-{\mathbf q}) P_c^{11} (k)  P_c^{11}(q), \nonumber \\
P_{c\nu}^{13} (k) &=& 3 \int d^3 {\mathbf q} \ F_3^{(s)}({\mathbf k},{\mathbf q},-{\mathbf q}) P_{c\nu}^{11} (k)  P_c^{11}(q),
\end{eqnarray}
where the subscript ``(s)'' denotes symmetrised (under the exchange of the wave-vectors in the argument) versions of the kernels given in equation~(\ref{eq:kernels}).
At this point the astute reader may question how a set of kernels originally derived 
for a pure CDM cosmology can be applied in this case of mixed CDM+neutrino cosmology,
especially given that  the linear propagator in this case is $k$-dependent.
Corrections to these expressions have in fact been derived for this very case of mixed CDM+neutrino cosmology~\cite{Wong:2008ws}.  However, for the same small neutrino masses required to validate the linear neutrino perturbation assumption, the corrections at $k \sim 0.1 \to 1\ {\rm Mpc}^{-1}$ are very small---generally no more than a percent.  In the same vein, the one-loop corrections~(\ref{eq:powerspectra}) are often also extended to $\Lambda$CDM 
cosmologies without further modification~\cite{Bernardeau:2001qr}.

Figure~\ref{fig:suppression} shows the one-loop corrected matter power spectrum for several massive neutrino cosmologies relative to the $\Lambda$CDM case.  
The main feature is an enhanced suppression on small scales due to neutrino free-streaming: the suppression is stronger than is predicted from linear perturbation theory, 
which is qualitatively consistent with the outcome of $N$-body simulations in figure~\ref{fig:damping}.

\begin{figure}[t]
\begin{center}
\includegraphics[width=16cm]{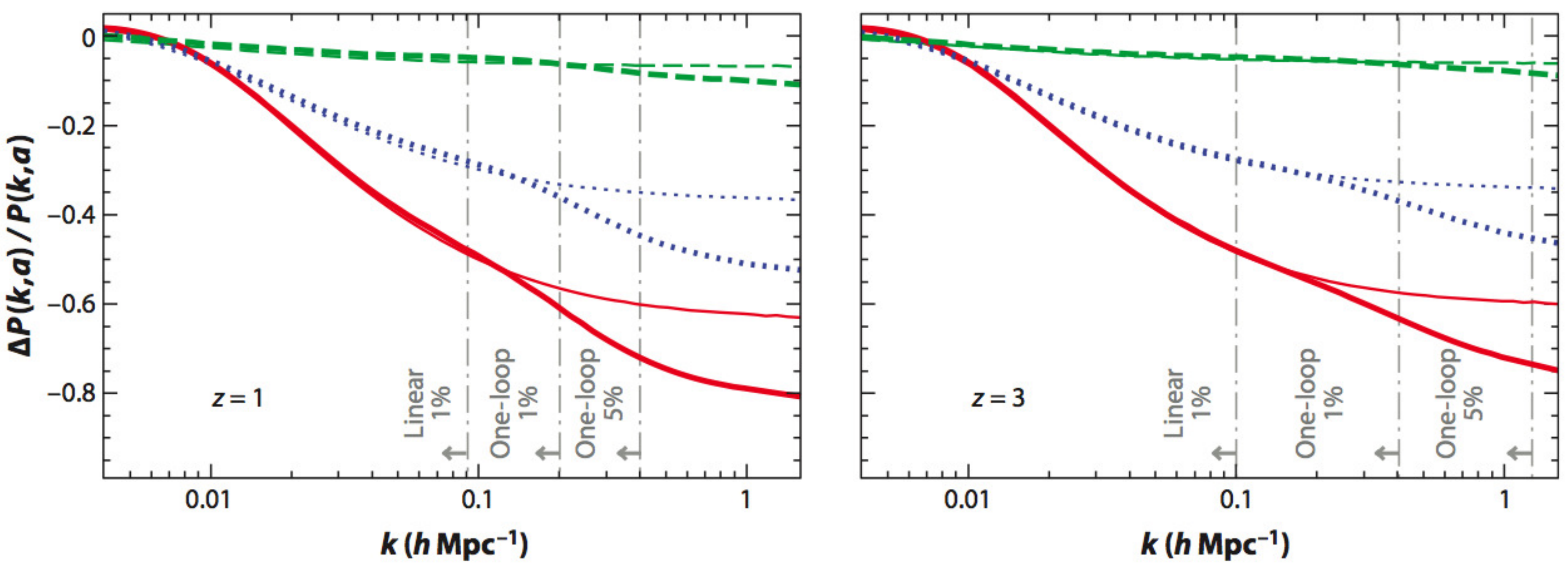}
\end{center}
\caption{Relative differences between the total matter power spectra
for a pure $\Lambda$CDM cosmology and  three models with massive neutrinos,
with $f_\nu=\Omega_\nu/\Omega_m=0.1$ (red/solid), 0.05 (blue/dotted), and 0.01 (green/dash)
at $z=1$ (left) and $z=3$ (right).  Thick lines indicate results 
including the one-loop correction, while the linear results 
are represented by the thin lines.  The three vertical lines indicate 
the maximum $k$ values at which the linear and the one-loop corrected matter
power spectra are accurate to better than 1\% and 5\%.\label{fig:suppression}}
\end{figure}

\subsection{Resummation/Renormalisation Group}

The time-flow renormalisation group (RG) method  is a simple extension of the perturbative method introduced above, but incorporates resummation 
of certain classes of perturbative terms to all orders~\cite{Pietroni:2008jx}.  Aside from its simplicity and ease-of-use, the time-flow RG method
has the advantage that it has been formulated for a broad class of cosmological models, including those with massive neutrinos.  
Other resummation approaches---of which there are many---are in their present formulations strictly speaking limited to $\Lambda$CDM or pure CDM type models.

Recall the master equation~(\ref{eq:compact}) for $\varphi$.  Ultimately, we are not interested in  the density contrast or velocity divergence   {\it per se}, 
but rather the $n$-point statistics such as the power spectrum
$\langle \varphi_a ({\mathbf k},s) \varphi_b({\mathbf q},s) \rangle = \delta_D ({\mathbf k}+{\mathbf q}) P_{ab} (k;s)$.  Therefore, it makes sense
 to construct directly equations of motion for the $n$-point correlators from the master equation~(\ref{eq:compact}).   For the two-point correlator,  we find
\begin{eqnarray}
\partial_s \langle \varphi_a \varphi_b \rangle\!\! \!\!&=& \!\!\!\! \langle \varphi_a \partial_\tau \varphi_b \rangle + \langle (\partial\tau \varphi_a) \varphi_b \rangle \\
\!\!\!\! &=&\!\!\!\! -\Pi_{bc}  \langle \varphi_a \varphi_c \rangle -\Pi_{ac}  \langle \varphi_b \varphi_c \rangle \!+\!\int \! d^3 q_1 d^3 q_2 (\gamma_{bcd}  \langle \varphi_a \varphi_c \varphi_d\rangle + \gamma_{acd}  \langle \varphi_b \varphi_c \varphi_d\rangle).\nonumber
\end{eqnarray}
Nonlinear coupling leads to a three-point correlator $ \langle \varphi_a \varphi_c \varphi_d\rangle$ 
to appear in the expression, which requires that we write out the corresponding equation of motion.  
However, nonlinear coupling necessarily introduces a four-point correlator into the equation of motion for the three-point correlator.
Thus a pattern emerges: the presence nonlinear coupling gives rise to an infinite hierarchy of differential equations for the $n$-point correlators, 
and this hierarchy must be truncated if we wish to produce a result.

In the time flow RG scheme, the hierarchy is truncated by simplifying the four-point correlator.
In general the four-point function is given by
\begin{equation}
\langle \varphi_a \varphi_b \varphi_c \varphi_d\rangle = \langle \varphi_a \varphi_b \rangle \langle \varphi_c \varphi_d\rangle + 
\langle \varphi_a \varphi_c \rangle \langle \varphi_b \varphi_d\rangle + \langle \varphi_a \varphi_d \rangle \langle \varphi_b \varphi_c\rangle +
 \langle \varphi_a \varphi_b \varphi_c \varphi_d\rangle_c,
\end{equation}
where we have implicitly assumed $\langle \varphi_a \rangle = 0$ (i.e., the mean density perturbation and mean velocity divergence are zero).
The connected piece, denoted by the subscript ``c'', can be written as
$\langle \varphi_a \varphi_b \varphi_c \varphi_d\rangle_c = \delta_D({\mathbf k}+{\mathbf q}+{\mathbf p}+{\mathbf r})Q_{abcd}({\mathbf k},{\mathbf q},{\mathbf p},{\mathbf r},s)$,
where $Q_{abcd}$ is the trispectrum. We set $Q_{abcd}$  to zero.

Thus, expressed in terms of the power spectrum $P_{ab}(k;s)$ and the bispectrum $B_{abc}({\mathbf k},{\mathbf q},{\mathbf r};s)$, where
$\langle \varphi_a ({\mathbf k},s) \varphi_b({\mathbf q},s) \varphi_c({\mathbf r},s) \rangle = \delta_D ({\mathbf k}+{\mathbf q}+{\mathbf r}) B_{abc} ({\mathbf k},{\mathbf q}, {\mathbf r};s)$, 
the final set of equations to be solved is~\cite{Pietroni:2008jx},
\begin{eqnarray}
\partial_s P_{ab} ({\mathbf k};s) &=& -\Pi_{ac} ({\mathbf k};s) P_{bc}({\mathbf k};s) - \Pi_{bc} ({\mathbf k};s) P_{ac}({\mathbf k};s) \nonumber \\
&& +\int d^3 {\mathbf q} \ [\gamma_{acd}({\mathbf k},-{\mathbf q},{\mathbf q}-{\mathbf k}) B_{bcd}({\mathbf k},-{\mathbf q},{\mathbf q}-{\mathbf k};s)
\nonumber \\
&& \hspace{15mm}+ \gamma_{bcd}({\mathbf k},-{\mathbf q},{\mathbf q}-{\mathbf k}) B_{acd}({\mathbf k},-{\mathbf q},{\mathbf q}-{\mathbf k};s)],\nonumber
\end{eqnarray}
\begin{eqnarray}
\partial_s B_{abc} ({\mathbf k},-{\mathbf q},{\mathbf q}-{\mathbf k};s) &=& -\Pi_{ad} ({\mathbf k};\tau) B_{dbc}({\mathbf k},-{\mathbf q},{\mathbf q}-{\mathbf k};s) \nonumber \\
&& \hspace{-25mm}-\Pi_{bd} ({\mathbf k};s) B_{adc}({\mathbf k},-{\mathbf q},{\mathbf q}-{\mathbf k};\tau) -\Pi_{cd} ({\mathbf k};\tau) B_{abd}({\mathbf k},-{\mathbf q},{\mathbf q}-{\mathbf k};s)\nonumber \\
&& \hspace{-25mm} + 2 [\gamma_{ade}({\mathbf k},-{\mathbf q},{\mathbf q}-{\mathbf k}) P_{bd}({\mathbf q};s) P_{ce}({\mathbf k}-{\mathbf q};s)\nonumber \\
&& \hspace{-15mm}+\gamma_{bde}(-{\mathbf q},{\mathbf q}-{\mathbf k},{\mathbf k}) P_{cd}({\mathbf k}-{\mathbf q};\tau) P_{ae}({\mathbf k};s)
\nonumber \\
&& \hspace{-15mm} +\gamma_{cde}({\mathbf q}-{\mathbf k},{\mathbf k},-{\mathbf q}) P_{ad}({\mathbf k};\tau) P_{be}({\mathbf q};s)].
\end{eqnarray}
Reference~\cite{Lesgourgues:2009am} applies these equations to the massive neutrino problem, again assuming that the neutrino perturbations remain at linear order,
 and demonstrate that they reproduce the absolute matter power spectra from the $N$-body simulations of  reference~\cite{Brandbyge:2008js} to a few percent accuracy
 up to $k \sim 1\  h \ {\rm Mpc}^{-1}$ at high redshifts ($z>2$).  This is  superior to the one-loop perturbative result, which has the same accuracy only up to  $k \sim 0.4\  h \ {\rm Mpc}^{-1}$
 at the same redshifts.  Interestingly, however, the time-flow RG and the one-loop perturbative approaches perform equally well when it comes to 
 predicting the relative suppression due to neutrino free-streaming at $z>2$.  Here, percent accuracy can be claimed up to $k \sim 1 \ h \ {\rm Mpc}^{-1}$.  See
 figures 2 and 3 of reference~\cite{Lesgourgues:2009am}.

\section{Future observations}\label{sec:future}

\subsection{Cosmic microwave background}

The Planck spacecraft, launched in May 2009, is expected to provide measurements of the
 CMB temperature fluctuations limited only by cosmic variance up to a multipole of 
$\ell \sim 1000$~\cite{planck}.  Cosmic variance-limited observations of the $E$-mode polarisation are expected to be possible 
up to $\ell$ of several hundred.  For the neutrino mass measurement, these improvements
imply that Planck alone will have a sensitivity to $\sum m_\nu$ as good as 
0.2~eV for the vanilla+$m_\nu$ model~(e.g., \cite{Perotto:2006rj}).%
\footnote{All forecasted sensitivities given in this section refer to the vanilla+$m_\nu$ model unless otherwise indicated.
Sensitivity is defined as the $68\%$ upper limit an observation can place on $\sum m_\nu$ if the true $\sum m_\nu$ is zero.} 
Even for more complicated model frameworks including a variable $N_{\rm eff}$ and dynamical dark energy,
we can anticipate $\sigma(\sum m_\nu) \sim 0.4$ from Planck alone~\cite{Perotto:2006rj}.
Importantly,  notwithstanding the improved sensitivities, linear perturbation theory will remain applicable
to CMB physics---no other observation will take its place as the most robust probe of cosmology in the near future.

\subsection{Large-scale structure}

Galaxy redshift surveys over the next decade will probe the LSS
at higher redshifts than before (up to $z \sim 4$).
The advantage is two-fold: firstly, nonlinear effects are less pronounced at
early times, meaning that it will be possible to observe up to $k \sim 1 \ {\rm Mpc}^{-1}$ 
with only mildly nonlinear corrections (however,  scale-dependent galaxy bias will still be an issue~\cite{Swanson:2010sk}).
 Secondly, going to higher redshifts will enable us to observe a larger 
volume $V$, thereby reducing the sampling uncertainty which scales as $\Delta P/P \sim (V k^3)^{-1/2}$.
The HETDEX, WFMOS, and BOSS  surveys
are envisaged to push $\sigma(\sum m_\nu)$ down to  0.1~eV when combined with Planck, 
while  with JDEM or EUCLID,  $\sigma(\sum m_\nu) \sim 0.05$~eV may be
possible~\cite{Lesgourgues:2004ps,Takada:2005si,Hannestad:2007cp,Abdalla:2007ut,Lahav:2009zr,Carbone:2010ik}.

Concerning Ly-$\alpha$,  barring uncertainties in the gas--bias relation, 
the Keck and/or VLT spectra of the Ly-$\alpha$  absorption lines have the potential to yield  $\sigma(\sum
m_{\nu})  \sim 0.05$~\cite{Gratton:2007tb} when combined with Planck.

Perhaps more interestingly, new kinds of  observations utilising different techniques to probe the large-scale structure 
distribution will also become available.   I describe some examples below.

\subsubsection{Weak gravitational lensing of galaxies} 

Light rays from distant galaxies are bent by 
matter density perturbations 
between the source galaxies and the observer, 
thereby inducing 
distortions in the observed images of the source galaxies.
By measuring the angular correlation of these distortions, one can 
probe the clustering statistics of the intervening matter density field~\cite{Bartelmann:1999yn}.

The current generation of weak lensing surveys are already producing interesting constraints on the neutrino mass (see section~\ref{sec:present}).
Future dedicated lensing surveys will probe higher redshifts (up to $z \sim 3$) with full sky coverage.
Furthermore, all these surveys will provide photometric redshift information on the source galaxies, allowing 
 for the binning of galaxy images by redshift and hence tomographic
studies of the evolution of the intervening large scale structure.
The Large Synoptic Survey Telescope (LSST)~\cite{lsst} combined with  Planck offers a sensitivity of $\sigma(\sum m_\nu) \sim 0.04$~eV using tomography.
Similar sensitivity is expected for DUNE/Euclid~\cite{Cooray:1999rv,Abazajian:2002ck,Song:2004tg,Hannestad:2006as,Kitching:2008dp,Namikawa:2010re}.

Dominant systematics include uncertainties in the photometric redshift measurements, which are crucial 
for tomographic studies.   Image distortions arising from the atmosphere and imperfections in the instrumentation must also be accurately known.
Because future lensing surveys will derive most of their constraining power at nominally nonlinear scales 
$k > 0.1 \ {\rm Mpc}^{-1}$,    uncertainties in our predictions of the 
nonlinear power spectrum must  be controlled to the percent level.  Baryon physics will also be important here.  The study 
of~\cite{Jing:2005gm} finds that baryon physics can contribute an uncertainty of up to 10\% at multipole $\ell > 1000$
 corresponding to physical scales $k$ of less than a factor of ten beyond the linear regime.

\subsubsection{Weak lensing of the CMB} 

Light from the last scattering surface can also 
be lensed by the foreground matter \cite{Lewis:2006fu}. 
 Indeed, with Planck
one can hope to probe $\sum m_\nu$ with sensitivity of $0.1 \ {\rm eV}$ via CMB lensing
\cite{Kaplinghat:2003bh,Lesgourgues:2005yv,Perotto:2006rj}.
The main advantage of CMB lensing over galaxy lensing is that because the source lies at such a high redshift ($z \sim 1000$),  the lenses probed are mainly those lying at $z \sim 4$, where the density perturbations are still evolving at  linear order.  The drawback, however, is that 
 tomographic studies are not possible because there is only one last scattering surface.
Another potential problem is that  extraction of the lensing signal hinges on the non-gaussian nature of the signal.  Other sources 
of non-gaussianity, whether primordial or secondary, may be an obstacle.

\subsubsection{Cluster mass function} 

As shown in figure~\ref{fig:cluster}, the halo mass function
is  sensitive to the absolute neutrino mass scale, because the formation 
of virialised objects is directly linked to the amplitude of the initial density perturbations.
The halo mass function is also known as the cluster mass function, since it is the observation 
of galaxy clusters that ultimately allows us to identify virialised objects in the low redshift 
universe.

As discussed in section~\ref{sec:present}, current X-ray observations of several hundred clusters have already 
produced  interesting 
limits on the neutrinos mass.  Future optical surveys such as the LSST are expected to yield cluster numbers in the thousands, thereby pushing to 
sensitivity down to  $\sigma (\sum m_{\nu}) = 0.04$~eV when 
their data are combined with data from Planck~\cite{Wang:2005vr}.
The main systematic challenge to these observations is how to determine the mass of the observed object.
For X-ray observations, for example, the X-ray temperature is a proxy for the cluster mass; the relation between the two
must be calibrated either through numerical simulations or through other observational means such as gravitational 
lensing effects.

A variant of these methods uses the Sunyaev--Zel'dovich effect to probe the cluster mass 
function~\cite{Shimon:2010gs}.   However,  the expected $\sigma(\sum m_\nu)$ is 
only $\sim 0.28$~eV.

\subsubsection{21 cm}  

Neutral hydrogen atoms can be found in two energy states depending on the alignment of the proton and the electron's spins.
A spin flip may occur taking the atom from the high to the low energy state by emission of a photon with a wavelength of 21cm.  Conversely,
absorption of a 21cm photon brings the atom from the low to the high energy  state.  Neutral hydrogen is the main constituent of the baryonic content 
of the universe and its distribution in space at high redshifts is expected to trace the underlying dark matter density field.  Thus, a measurement 
of the spatial fluctuations in the 21cm brightness temperature $T_b$ could in principle be a powerful way to map  the LSS distribution~\cite{Furlanetto:2006jb}.

Current and upcoming radio arrays such as LOFAR, MWA, and SKA are designed primarily to study the physics of reionisation at redshifts $6<z< 12$ when neutral hydrogen 
reside in the intergalactic medium.
This is less than ideal for the purpose of mapping the LSS distribution, since fluctuations in $T_b$ in this
epoch depend not only on the matter density perturbations, but 
also on the process of reionisation, which is probably a spatially inhomogeneous process.   If reionisation physics 
can be accurately modelled, then it should be possible to 
reach $\sigma(\sum m_\nu) \sim 0.02$~eV using a combination of Planck and
 SKA~\cite{Mao:2008ug,Pritchard:2008wy}.   Gravitational lensing of the 21 cm signal 
could also yield similar sensitivities~\cite{metcalf1,metcalf2}.
Innovative experimental design such as the proposed 
Fast Fourier Transform Telescope (FFTT)~\cite{fftt} could even yield $0.003$~eV, 
which would enable one to probe even the individual neutrino 
mass~\cite{Mao:2008ug,Pritchard:2008wy}.  However, if no accurate modelling of reionisation is available, 
then the sensitivities to $\sum m_\nu$ would significantly degrade, yielding
$\sigma(\sum{m_\nu})\sim0.02$~eV from Planck+FFTT~\cite{Mao:2008ug}.   

Note that, with present technology, it is not possible to observe at higher redshifts to skirt the issue of reionisation. 
From the perspective of cosmological parameter estimation, 
observations at lower redshifts when neutral hydrogen forms dense clumps in galaxies 
offer no real advantage over conventional galaxy redshift surveys.

\section{Conclusions\label{sec:conclusions}}

Precision cosmology offers an interesting way to probe many aspects of fundamental physics.
Neutrino properties are an excellent example.  By using probes of the large-scale structure and 
of the cosmic microwave background anisotropies, we can already deduce that the sum of the 
neutrino masses must be less than about $1$~eV regardless of the precise details of the cosmological model.
 This limit is already  better than
those from laboratory experiments.
Future cosmological observations will perform even better: in the most optimistic case, a
sensitivity of  $\sigma(\sum_\nu) \sim 0.04 \ {\rm eV}$ may be achievable
with a combination of Planck, galaxy redshift surveys, weak gravitational lensing observations, and so on.

From a theoretical perspective,  the key issue is to maintain control over the nonlinear physics: 
specifically nonlinear evolution 
of the density perturbations and the galaxy bias.  To this end, 
there have been recent developments in the computation of the nonlinear power spectrum and other observables to percent level accuracy
by numerical and semi-analytical means.  Understanding the nonlinear issue will allow us to use future observations to their maximum
potential, and hopefully to realise the goal of measuring the absolute neutrino mass scale with precision cosmology one day.



\begin{thebibliography}{99}

\bibitem{nuexp}
 Gonzalez-Garcia MC, Maltoni M, Salvado J. 2010.
 Updated global fit to three neutrino mixing: status of the hints of theta13 $>$ 0.
{\it  JHEP} 1004:056

\bibitem{bbn}
Iocco F et al. 2009.
Primordial Nucleosynthesis: from precision cosmology to fundamental physics.
{\it Phys. Rept.} 472:1--76

\bibitem{Lobashev:2003kt}
Lobashev VM. 2003.
The search for the neutrino mass by direct method in the tritium beta-decay
and perspectives of study it in the project KATRIN. 
{\it Nucl. Phys.} A719:153--160

\bibitem{Kraus:2004zw}
  Kraus C et al. 2005.
  Final results from phase II of the Mainz neutrino mass search in  tritium
  beta decay.
{\it   Eur. Phys. J.} C40:447--468


\bibitem{Lesgourgues:2006nd}
Lesgourgues J, Pastor S. 2006.
Massive neutrinos and cosmology.
{\it Phys. Rept.}  429:307--379

\bibitem{neff}
  Mangano G, Miele G, Pastor S et al. 2005.
  Relic neutrino decoupling including flavor oscillations.
{\it  Nucl. Phys.}  B729:221--234


\bibitem{Fixsen:1996nj}
Fixsen DJ et al. 1996.
  The Cosmic Microwave Background Spectrum from the Full COBE/FIRAS Data Set.
{\it  Astrophys.  J.}  473:576


\bibitem{Gershtein:1966gg}
Gershtein SS, Zeldovich YB. 1966.
Rest mass of muonic neutrino and cosmology.
{\it JETP Lett.}  4:120--122

\bibitem{Cowsik:1972gh}
Cowsik R, McClelland J. 1972.
An Upper Limit on the Neutrino Rest Mass.
{\it  Phys. Rev. Lett.}  29:669--670

 \bibitem{tremainegunn}
  Tremaine S, Gunn JE. 1979.
 Dynamical Role of Light Neutral Leptons in Cosmology.
{\it  Phys. Rev. Lett.}   42:407--410
  
 \bibitem{mabertschinger}
  Ma CP, Bertschinger E. 1995.
  Cosmological perturbation theory in the synchronous and conformal Newtonian gauges.
{\it  Astrophys.  J.}  455:7--25



\bibitem{durrer}
Durrer, R. 2009.
The Cosmic Microwave Background.
Cambridge University Press. 424 pp.




\bibitem{Komatsu:2010fb}
 Komatsu E et al. 2010.
  Seven-Year Wilkinson Microwave Anisotropy Probe (WMAP) Observations:
  Cosmological Interpretation.
  arXiv:1001.4538 [astro-ph.CO]

\bibitem{riotto}
  Lyth DH, Riotto A. 1999.
 Particle physics models of inflation and the cosmological density perturbation.
 {\it  Phys.  Rept.}   314:1-146

\bibitem{Lewis:1999bs}
  Lewis A, Challinor A, Lasenby A. 2000.
  Efficient computation of CMB anisotropies in closed FRW models.
{\it  Astrophys.  J.}   538:473--476.



\bibitem{Jarosik:2010iu}
  Jarosik N et al. 2010.
  Seven-Year Wilkinson Microwave Anisotropy Probe (WMAP) Observations: Sky
  Maps, Systematic Errors, and Basic Results.
  arXiv:1001.4744 [astro-ph.CO]

\bibitem{Reichardt:2008ay}
  Reichardt CL et al. 2009.
  High resolution CMB power spectrum from the complete ACBAR data set.
 {\it  Astrophys. J.}  694:1200--1219



\bibitem{Chiang:2009xsa}
  Chiang HC et al. 2010.
  Measurement of CMB Polarization Power Spectra from Two Years of BICEP
  Data.
{\it Astrophys. J.} 711:1123--1140

\bibitem{Brown:2009uy}
  Brown ML et al. [QUaD collaboration]. 2009.
  Improved measurements of the temperature and polarization of the CMB from
  QUaD.
{\it  Astrophys. J.}  705:978--999


\bibitem{Das:2010ga}
Das S et al. 2010.
  The Atacama Cosmology Telescope: A Measurement of the Cosmic Microwave Background Power Spectrum at 148 and 218 GHz from the 2008 Southern Survey.
 arXiv:1009.0847 [astro-ph.CO]


\bibitem{Hinshaw:2006ia}
 Hinshaw G et al. [WMAP Collaboration]. 2007.
  {\it Astrophys. J. Suppl.}  170:288


\bibitem{het}
  Hu W, Eisenstein DJ, Tegmark M. 1998.
  Weighing neutrinos with galaxy surveys.
{\it  Phys. Rev. Lett.}   80:5255--5258



\bibitem{Ringwald:2004np}
  Ringwald A, Wong YYY. 2004.
  Gravitational clustering of relic neutrinos and implications for their detection.
{\it  J. Cosmo. Astropart. Phys.}  0412:005


\bibitem{Cole:2005sx}
  Cole S et al. [The 2dFGRS Collaboration]. 2005.
  The 2dF Galaxy Redshift Survey: Power-spectrum analysis of the final
  dataset and cosmological implications.
 {\it  Mon. Not. Roy. Astron. Soc.}  362:505--534



\bibitem{Abazajian:2008wr}
  Abazajian KN et al.  [SDSS Collaboration]. 2009.
  The Seventh Data Release of the Sloan Digital Sky Survey.
  {\it Astrophys. J. Suppl.}  182:543--558




\bibitem{tegmarklrg}
  Tegmark M et al. [ SDSS Collaboration ]. 2006.
 Cosmological Constraints from the SDSS Luminous Red Galaxies.
  {\it Phys.  Rev.}   D74:123507
  



\bibitem{Reid:2009xm}
  Reid BA et al. 2010.
  Cosmological Constraints from the Clustering of the Sloan Digital Sky Survey DR7 Luminous Red Galaxies.
  {\it Mon. Not. Roy. Astron. Soc.}  404:60--85

\bibitem{bias}
  Hamann J, Hannestad S, Melchiorri A, Wong, YYY. 2008.
 Nonlinear corrections to the cosmological matter power spectrum and scale-dependent galaxy bias: Implications for parameter estimation.
 {\it  J. Cosmo. Astropart. Phys.}  0807:017



\bibitem{Hamann:2010pw}
  Hamann J et al. 2010.
  Cosmological parameters from large scale structure - geometric versus shape information.
  {\it J. Cosmo. Astropart. Phys.} 1007:022





\bibitem{McDonald:2004xn}
  McDonald P  et al. [ SDSS Collaboration ]. 2005.
 The Linear theory power spectrum from the Lyman-alpha forest in the Sloan Digital Sky Survey.
{\it  Astrophys.  J.}   635:761--783


\bibitem{Viel:2005eg}
  Viel M, Haehnelt MG, Springel V. 2006.
  Testing the accuracy of the hydro-PM approximation in numerical simulations of the Lyman-alpha forest.
  {\it Mon. Not. Roy. Astron. Soc.}  367:1655--1665


\bibitem{Viel:2005ha}
  Viel M, Haehnelt MG. 2006.
  Cosmological and astrophysical parameters from the SDSS flux power spectrum and hydrodynamical simulations of the Lyman-alpha forest.
{\it  Mon. Not. Roy. Astron. Soc.}  365:231--244


\bibitem{Viel:2006yh}
  Viel M, Haehnelt MG, Lewis . 2006.
 The Lyman-alpha forest and WMAP year three.
{\it  Mon. Not. Roy. Astron. Soc.}  370:L51--L55


\bibitem{Percival:2009xn}
  Percival WJ et al. 2010.
  Baryon Acoustic Oscillations in the Sloan Digital Sky Survey Data Release 7
  Galaxy Sample.
{\it  Mon. Not. Roy. Astron. Soc.}  401:2148--2168

  
\bibitem{Riess:2009pu}
  Riess AG et al. 2009.
  A Redetermination of the Hubble Constant with the Hubble Space Telescope
  from a Differential Distance Ladder.
 {\it Astrophys. J.}  699:539--563
  

\bibitem{Hannestad:2010yi}
  Hannestad S, Mirizzi A, Raffelt GG, Wong YYY. 2010.
 Neutrino and axion hot dark matter bounds after WMAP-7.
{\it  J. Cosmo. Astropart. Phys.}  1008:001

\bibitem{Reid:2009nq}
  Reid BA, Verde L, Jimenez R  et al. 2010.
 Robust Neutrino Constraints by Combining Low Redshift Observations with the CMB.
{\it  J. Cosmo. Astropart. Phys.} 1001:003







\bibitem{Thomas:2009ae}
  Thomas SA, Abdalla FB, Lahav O. 2010.
  Upper Bound of 0.28eV on the Neutrino Masses from the Largest Photometric Redshift Survey.
  {\it Phys. Rev. Lett.}  105:031301

\bibitem{astier}
Astier, P et al. 2006.
The Supernova legacy survey: Measurement of omega(m), omega(lambda) and W from the first year data set.
{\it Astron. Astrophys.} 447:31--48





\bibitem{sn}
Kessler R et al. 2009.
First Sloan Digital Sky Survey-II (SDSS-II) Supernova Results: Hubble Diagram and Cosmological Parameters.
{\it Astrophys. J. Suppl.} 185:32--84 



\bibitem{Ichiki:2008ye}
  Ichik K, Takada M, Takahashi T. 2009.
 Constraints on Neutrino Masses from Weak Lensing.
{\it  Phys. Rev.}   D79:023520


\bibitem{kristiansen}
 Kristiansen JR, Elgar{\o}y {\O}, Dahle H. 2007.
Using the cluster mass function from weak lensing to constrain neutrino masses.
{\it Phys. Rev.} D75:083510


\bibitem{Mantz:2009rj}
  Mantz A, Allen SW, Rapetti D. 2010.
  The Observed Growth of Massive Galaxy Clusters IV: Robust Constraints on Neutrino Properties.
{\it  Mon. Not. Roy. Astron. Soc.}  406:1805--1814



\bibitem{Zunckel:2006mt}
  Zunckel C, Ferreira PG. 2007.
  Conservative Estimates of the Mass of the Neutrino from Cosmology.
{\it  J. Cosmo. Astropart. Phys.}  0708:004


\bibitem{copeland}
Copeland EJ, Sami M, Tsujikawa S. 2006.
Dynamics of dark energy.
{\it Int. J. Mod. Phys.} D15:1753--1936 


\bibitem{LaVacca:2009yp}
  La Vacca G, Kristiansen JR, Colombo LPL et al. 2009.
  Do WMAP data favor neutrino mass and a coupling between Cold Dark Matter and Dark Energy?
{\it  J. Cosmo. Astropart. Phys.} 0904:007



\bibitem{Hannestad:2005gj}
  Hannestad S. 2005.
 Neutrino masses and the dark energy equation of state: Relaxing the
  cosmological neutrino mass bound.
{\it   Phys. Rev. Lett.}  95:221301



\bibitem{starobinsky}
Motohashi, H, Starobinsky, AA, Yokoyama, J. 2010.
Matter power spectrum in f(R) gravity with massive neutrinos.
{\it Prog. Theor. Phys.} 123:541--546 


\bibitem{GonzalezGarcia:2010un}
  Gonzalez-Garcia MC, Maltoni M, Salvado J. (2010)
  Robust Cosmological Bounds on Neutrinos and their Combination with Oscillation Results.
{\it  JHEP} 1008:117
  

\bibitem{Blanchard:2003du}
  Blanchard A, Douspis M, Rowan-Robinson M, Sarkar S. 2003.
  An alternative to the cosmological 'concordance model'.
{\it  Astron.  Astrophys.}   412:35--44



\bibitem{Hunt:2007dn}
  Hunt P, Sarkar S. 2007.
 Multiple inflation and the WMAP glitches. 2. Data analysis and cosmological parameter extraction.
 {\it Phys. Rev.}  D76:123504


\bibitem{katrin}
Th\"ummler, T [KATRIN Collaboration]. 2010.
Introduction to direct neutrino mass measurements at KATRIN.
arXiv:1012.2282 [hep-ex]

\bibitem{Bertschinger:1998tv}
  Bertschinger, E. 1998.
  Simulations of structure formation in the universe.
{\it  Ann. Rev. Astron. Astrophys.}  36:599--654



\bibitem{Jing:2005gm}
  Jing YP, Zhang P, Lin WP  et al. 2006.
  The influence of baryons on the clustering of matter and weak lensing surveys.
{\it  Astrophys.  J.}  640:L119--L122.


\bibitem{Barnes:1986nb}
  Barnes J, Hut P. 1986.
  A Hierarchical 0 Nlogn Force Calculation Algorithm.
{\it  Nature} 324:446--449


\bibitem{gadget}
  Springel, V. 2005.
 The Cosmological simulation code GADGET-2.
 {\it  Mon. Not. Roy. Astron. Soc.}  364:1105--1134




\bibitem{Brandbyge:2008rv}
  Brandbyge J, Hannestad S, Haugb{\o}lle T, Thomsen, B.  2008.
  The Effect of Thermal Neutrino Motion on the Non-linear Cosmological Matter Power Spectrum.
  {\it J. Cosmo. Astropart. Phys.} 0808:020
  
\bibitem{Viel:2010bn}
  Viel M, Haehnelt MG, Springel V. 2010.
 The effect of neutrinos on the matter distribution as probed by the Intergalactic Medium.
{\it  J. Cosmo. Astropart. Phys.} 1006:015




\bibitem{Brandbyge:2008js}
  Brandbyge J, Hannestad S. 2009.
  Grid Based Linear Neutrino Perturbations in Cosmological N-body Simulations.
  {\it J. Cosmo. Astropart. Phys.} 0905:002


\bibitem{Brandbyge:2009ce}
  Brandbyge J, Hannestad S. 2010.
  Resolving Cosmic Neutrino Structure: A Hybrid Neutrino N-body Scheme.
  {\it J. Cosmo. Astropart. Phys.} 1001:021


\bibitem{Brandbyge:2010ge}
  Brandbyge J, Hannestad S, Haugb{\o}lle T, Wong, YYY.  2010.
  Neutrinos in Non-linear Structure Formation - The Effect on Halo Properties.
  {\it J. Cosmo. Astropart. Phys.} 1009:014






\bibitem{Bernardeau:2001qr}
  Bernardeau F, Colombi S, Gaztanaga E et al. 2002.
 Large scale structure of the universe and cosmological perturbation theory.
{\it  Phys. Rept.}  367:1--248


\bibitem{Saito:2008bp}
Saito S, Takada M, Taruya A. 2008.
  Impact of massive neutrinos on nonlinear matter power spectrum.
{\it  Phys. Rev. Lett.}  100:191301


\bibitem{Wong:2008ws}
Wong YYY. 2008.
Higher order corrections to the large scale matter power spectrum in the presence of massive neutrinos.
{\it J. Cosmo. Astropart. Phys.}  0810:035


\bibitem{Shoji:2009gg}
Shoji M, Komatsu E. 2009.
Third-order Perturbation Theory With Non-linear Pressure.
{\it  Astrophys. J.}  700:705--719

\bibitem{Pietroni:2008jx}
  Pietroni, M. 2008.
 Flowing with Time: a New Approach to Nonlinear Cosmological Perturbations.
 {\it J. Cosmo. Astropart. Phys.}   0810:036



\bibitem{Lesgourgues:2009am}
  Lesgourgues J. Matarrese S, Pietroni M et al. 2009.
  Non-linear Power Spectrum including Massive Neutrinos: the Time-RG Flow Approach.
 {\it J. Cosmo. Astropart. Phys.} 0906:017



\bibitem{planck}
[Planck Collaboration] 2006.
The scientific programme of Planck.
astro-ph/0604069

\bibitem{Perotto:2006rj}
  Perotto L, Lesgourgues J, Hannestad S et al. 2006.
  Probing cosmological parameters with the CMB: Forecasts from full Monte
  Carlo simulations.
 {\it  J. Cosmo. Astropart. Phys.} 0610:013
  
  

\bibitem{Swanson:2010sk}
  Swanson MEC, Percival WJ, Lahav O. 2010.
 Neutrino masses from clustering of red and blue galaxies: a test of astrophysical uncertainties.
 {\it  Mon. Not. Roy. Astron. Soc.}   409:1100--1112
 

  
  
\bibitem{Lesgourgues:2004ps}
  Lesgourgues J, Pastor S, Perotto L. 2004.
  Probing neutrino masses with future galaxy redshift surveys.
{\it  Phys. Rev.}  D70:045016

\bibitem{Takada:2005si}
  Takada M, Komatsu E, Futamase T. 2006.
  Cosmology with high-redshift galaxy survey: Neutrino mass and  inflation.
 {\it Phys. Rev.} D73:083520


\bibitem{Hannestad:2007cp}
  Hannestad S, Wong YYY. 2007.
  Neutrino mass from future high redshift galaxy surveys: Sensitivity and
  detection threshold.
{\it  J. Cosmo. Astropart. Phys.} 0707:004
  
\bibitem{Abdalla:2007ut}
  Abdalla FB, Rawlings S. 2007.
  Determining neutrino properties using future galaxy redshift surveys.
{\it  Mon. Not. Roy.  Astron. Soc.} 381:1313--1328



\bibitem{Lahav:2009zr}
  Lahav O, Kiakotou A, Abdalla  FB et al. 2009.
 Forecasting neutrino masses from galaxy clustering in the Dark Energy Survey combined with the Planck Measurements.
  arXiv:0910.4714 [astro-ph.CO]



\bibitem{Carbone:2010ik}
  Carbone C, Verde L, Wang Y et al. 2010.
  Neutrino constraints from future nearly all-sky spectroscopic galaxy surveys.
   arXiv:1012.2868 [astro-ph.CO]








 
 
 
\bibitem{Gratton:2007tb}
  Gratton S, Lewis A, Efstathiou G. 2008.
  Prospects for Constraining Neutrino Mass Using Planck and Lyman-Alpha Forest Data.
{\it  Phys. Rev.}   D77:083507



\bibitem{Bartelmann:1999yn}
  Bartelmann M, Schneider P. 2001.
  Weak Gravitational Lensing.
{\it   Phys. Rept.}  340:291--472
  
  \bibitem{lsst}
  [LSST Science and LSST Project Collaborations] 2009.
 LSST Science Book, Version 2.0.
  arXiv:0912.0201 [astro-ph.IM]

  
  
  
\bibitem{Cooray:1999rv}
  Cooray AR. 1999.
  Weighing neutrinos: Weak lensing approach.
 {\it  Astron.  Astrophys.}  348:31

  
 
\bibitem{Abazajian:2002ck}
  Abazajian KN, Dodelson S. 2003.
 Neutrino mass and dark energy from weak lensing.
{\it  Phys. Rev. Lett.}  91:041301
 
\bibitem{Song:2004tg}
  Song YS, Knox L. 2004.
  Determination of cosmological parameters from cosmic shear data.
{\it  Phys.  Rev.}   D70:063510



\bibitem{Hannestad:2006as}
  Hannestad S, Tu H, Wong YYY. 2006.
  Measuring neutrino masses and dark energy with weak lensing tomography.
{\it   J. Cosmo. Astropart. Phys.} 0606:025

\bibitem{Kitching:2008dp}
  Kitching TD et al. 2008.
  Finding Evidence for Massive Neutrinos using 3D Weak Lensing.
  {\it Phys. Rev.}  D77:103008




\bibitem{Namikawa:2010re}
  Namikawa T, Saito S, Atsushi T. 2010.
  Probing dark energy and neutrino mass from upcoming lensing experiments of CMB and galaxies.
  arXiv:1009.3204 [astro-ph.CO]




\bibitem{Lewis:2006fu}
  Lewis A, Challinor A. 2006.
  Weak Gravitational Lensing of the CMB.
{\it Phys. Rept.}  429:1--65

\bibitem{Kaplinghat:2003bh}
  Kaplinghat M, Knox L, Song YS. 2003.
  Determining neutrino mass from the CMB alone.
  {\it Phys. Rev. Lett.} 91:241301

\bibitem{Lesgourgues:2005yv}
  Lesgourgues J, Perotto L, Pastor S et al. 2006.
  Probing neutrino masses with CMB lensing extraction.
  {\it Phys. Rev.}  D73:045021



\bibitem{Wang:2005vr}
 Wang S et al. 2005.
  Weighing neutrinos with galaxy cluster surveys.
 {\it Phys. Rev. Lett.}  95:011302



\bibitem{Shimon:2010gs}
  Shimon M, Sadeh S, Rephaeli Y. 2010.
  Neutrino Mass Inference from SZ Surveys.
  arXiv:1009.4110 [astro-ph.CO].

\bibitem{Furlanetto:2006jb}
  Furlanetto S,  Oh SP,  Briggs F. 2006.
  Cosmology at Low Frequencies: The 21 cm Transition and the High-Redshift
  Universe. 
{\it  Phys. Rept.}  433:181--301


  
\bibitem{Mao:2008ug}
  Mao Y, Tegmark M, McQuinn M  et al. 2008.
  How accurately can 21 cm tomography constrain cosmology?
{\it  Phys. Rev.}   D78:023529

\bibitem{Pritchard:2008wy}
  Pritchard JR, Pierpaoli E. 2008.
  Constraining massive neutrinos using cosmological 21 cm observations.
 {\it  Phys. Rev.}   D78:065009


\bibitem{metcalf1}
Metcalf RB, White SDM. 2009.
Cosmological Information in the Gravitational Lensing of Pregalactic HI.
{\it Mon. Not. Roy. Astron. Soc.} 394:704--714

\bibitem{metcalf2}
Metcalf RB.  2010.
Neutrino masses, dark energy and the gravitational lensing of pre-galactic HI.
{\it Mon. Not. Roy. Astron. Soc.} 401:1999--2004.


  
 \bibitem{fftt}
  Tegmark M, Zaldarriaga M. 2009.
 The Fast Fourier Transform Telescope.
 {\it Phys. Rev.}   D79:083530


\end{thebibliography}
\end{document}